%% file: paper.tex
\documentclass[conference,10pt,reprint]{sigplanconf}

\input{lake/preamble}

\usepackage{paralist}
\usepackage{booktabs}
\usepackage[sort&compress]{natbib}

\usepackage[normalem]{ulem}	
\usepackage[T1]{fontenc}
\usepackage{textcomp}
\usepackage{mathptmx}
\usepackage{fixltx2e}
\usepackage{stmaryrd}
\usepackage{caption}
\usepackage{subcaption}
\usepackage{balance}
\usepackage{beramono}
\usepackage{lipsum}
\DeclareMathAlphabet{\mathcal}{OMS}{cmsy}{m}{n} 

\usepackage{comment}
\usepackage{framed}

\usepackage{tabularx}
\usepackage{graphicx}

\newcolumntype{a}{>{$}c<{$}}
\newcolumntype{b}{>{$}l<{$}}
\newcolumntype{d}{>{$}r<{$}}

\newcommand{\ourtitle}{On the Lexical Distinguishability of Source Code}
\newcommand{\HS}{\textsc{Hitting-Set}\xspace}
\newcommand{\Minset}{\textsc{Minset}\xspace}

\newcommand{\Lex}{\textsc{LEX}\xspace}
\newcommand{\LTT}{\textsc{LTT}\xspace}
\newcommand{\MinOne}{\textsc{Min1}\xspace}
\newcommand{\MinTwo}{\textsc{Min2}\xspace}
\newcommand{\MinThree}{\textsc{Min3}\xspace}
\newcommand{\MinFour}{\textsc{Min4}\xspace}
\newcommand{\Noun}{\textsc{Noun}\xspace}
\newcommand{\Verb}{\textsc{Verb}\xspace}
\newcommand{\Adj}{\textsc{Adjective}\xspace}

\newcommand{\ID}{\textsc{Identifier}\xspace}
\newcommand{\INTLIT}{\textsc{IntLit}\xspace}

\newcounter{RQCounter}

\newcommand{\vs}{vs.\xspace}

\makeatletter
\newcounter{ALC@tempcntr}
\newcommand{\LCOMMENT}[1]{%
    \setcounter{ALC@tempcntr}{\arabic{ALC@rem}}
    \setcounter{ALC@rem}{1}
    \item \textit{// #1}
    \setcounter{ALC@rem}{\arabic{ALC@tempcntr}}
}%
\makeatother

\newlength{\emstr}
\newcommand{\boldpara}[1]{%
\smallskip%
\par\noindent\textbf{\textit{#1}}\hspace{\emstr}
}%

\makeatletter
\def\@copyrightspace{\relax}
\makeatother

\definecolor{javared}{rgb}{0.6,0,0} 
\definecolor{javagreen}{rgb}{0.25,0.5,0.35} 
\definecolor{javapurple}{rgb}{0.5,0,0.35} 
\definecolor{javadocblue}{rgb}{0.25,0.35,0.75} 

\definecolor{added-color}{RGB}{0, 75, 0}
\definecolor{removed-color}{RGB}{150, 0, 0}
\definecolor{unchanged-color}{RGB}{115, 115, 115}

\hypersetup{
pdftitle={\ourtitle},
    pdfauthor={Martin Velez, Dong Qiu, You Zhou, Earl T. Barr, Zhendong Su},
    plainpages=false,
	linkcolor=black, 
	citecolor=black, 
	filecolor=black,
	urlcolor=black,
    pdfpagelabels
}

\newcommand{\defref}[1]{\hyperref[#1]{Definition~\ref*{#1}}}
\newcommand{\algref}[1]{\hyperref[#1]{Algorithm~\ref*{#1}}}
\newcommand{\lineref}[1]{\hyperref[#1]{Line~\ref*{#1}}}

\title{\ourtitle%
}

\newcommand\Mark[1]{\textsuperscript#1}

\authorinfo{Martin Velez\Mark{1}\qquad Dong Qiu\Mark{2}\qquad You Zhou\Mark{1}\qquad Earl Barr\Mark{3}\qquad Zhendong Su\Mark{1}}
{\Mark{1}Department of Computer Science, University of California, Davis, USA\\
\Mark{2}School of Computer Science and Engineering, Southeast University, China\\
\Mark{3}Department of Computer Science, University College London, UK}
{\{marvelez,yyzhou,su\}@ucdavis.edu, dongqiu@seu.edu.cn, e.barr@ucl.ac.uk}

\begin{document}

\maketitle

\input{abstract}
\input{intro}

\input{threshing_and_winnowing}
\input{methodology}

\input{results}

\input{discussion}

\input{applications}

\input{relwork}

\input{conc}



\setlength{\bibsep}{.149cm} 
\bibliographystyle{abbrv}
\footnotesize
\balance
\bibliography{Bib/paper}

\end{document}

%% file: lake/preamble.tex
\usepackage{graphicx,color}
\usepackage{amsmath}
\usepackage{amsfonts}
\usepackage{amssymb}
\usepackage{amsthm}
\usepackage{url} 
\usepackage{multicol}
\usepackage{multirow}
\usepackage{setspace} 
\usepackage{xspace}
\usepackage{listings}
\usepackage{ifthen}
\usepackage{verbatim}
\usepackage{float} 
\usepackage{microtype}
\usepackage{mathptmx}
\usepackage{stmaryrd} 
\usepackage{textcomp}
\usepackage[sort&compress]{natbib} 
\usepackage{fixltx2e} 
\usepackage{booktabs}
\usepackage{fancyhdr}

\usepackage[mathletters]{ucs}
\usepackage[utf8x]{inputenc}
\usepackage{wasysym}	

\usepackage[bookmarks, colorlinks, citecolor=green, urlcolor=blue,
			filecolor=blue, linkcolor=blue]{hyperref}
\usepackage[noend]{algorithmic} 
\usepackage{algorithm}

\newcommand{\etal}{\hbox{\emph{et al.}}\xspace}
\newcommand{\eg}{\hbox{\emph{e.g.}}\xspace}
\newcommand{\ie}{\hbox{\emph{i.e.}}\xspace}

\newfloat{Protocol}{thp}{lop}
\newfloat{Program}{thp}{lop}
\newfloat{Procedure}{thp}{lop}

\newenvironment{proof-idea}{\noindent{\bf Proof Idea}\hspace*{1em}}{\bigskip}

\theoremstyle{plain}
\newtheorem{thm}{Theorem}[section]

\theoremstyle{remark}

\theoremstyle{definition}
\newtheorem{defn}{Definition}[section]

\DeclareUnicodeCharacter{183}{\cdot}						
\DeclareUnicodeCharacter{931}{\ensuremath\Sigma}			
\DeclareUnicodeCharacter{9001}{\ensuremath\langle}			
\DeclareUnicodeCharacter{9002}{\ensuremath\rangle}			
\DeclareUnicodeCharacter{9608}{\ensuremath\blacksquare}		
\DeclareUnicodeCharacter{1013}{\in}							
\DeclareUnicodeCharacter{8213}{---}							



%% file: abstract.tex
\begin{abstract}

Natural language is robust against noise.  The meaning of many sentences
survives the loss of words, sometimes many of them.  Some words in a sentence,
however, cannot be lost without changing the meaning of the sentence.  We call
these words ``wheat’’ and the rest ``chaff’’.  The word “not” in the sentence “I
do not like rain” is wheat and “do” is chaff.  For human understanding of the
purpose and behavior of \emph{source code}, we hypothesize that the same holds.
To quantify the extent to which we can separate code into ``wheat'' and
``chaff'', we study a large (100M LOC), diverse corpus of real-world projects in
Java.  Since methods represent natural, likely distinct units of code, we use
the $\sim$9M Java methods in the corpus to approximate a universe of
``sentences.''  We extract their wheat by computing the function's \emph{minimal
distinguishing subset (\Minset)}.  Our results confirm that functions contain
much chaff.  On average, \textsc{minsets} have 1.56 words (none exceeds 6) and
comprise 4\% of their methods.  Beyond its intrinsic scientific interest, our
work offers the first quantitative evidence for recent promising work on
keyword-based programming and insight into how to develop a powerful, alternative
programming model. 
 
\end{abstract}

%% file: intro.tex
\section{Introduction}
\label{sec:intro}

A basic but strong assumption underlies many research and engineering efforts
like code search, code completion, keyword programming, and natural programming:
From a ``small'' subset of words, a system can find or generate a larger,
executable piece of code.

This assumption is crucial in code search work.  
The body of work breaks the search problem into three subproblems
\begin{inparaenum}[1)] \item how to store and index
code~\cite{bajracharya2006oopsla,mcmillan2011icse}, \item what queries (and
results) to support~\cite{reiss2009icse,reiss2009suite}, and \item how to filter
and rank the
results~\cite{bajracharya2006oopsla,mandelin2005pldi,mcmillan2012icse}.
\end{inparaenum}  The person doing the search only has one concern: ``\emph{What
should I type to find the code I want?}''. 
Efforts focus on building
better search engines not on determining to what extent this assumption holds.

This assumption is also critical in keyword and natural programming
implementations~\cite{little_ase_07,little_nocode_10,le_mobisys_13,miller_uist_08}.
Almost a decade ago, Little \etal devised a keyword programming technique to
translate keyword queries into valid Java expressions~\cite{little_ase_07}.
Several tools and tools and techniques grouped under the general term of Sloppy
Programming followed ~\cite{little_nocode_10,miller_uist_08}.  These tools
interpret keyword queries directly by first translating them into source code.
SmartSynth~\cite{le_mobisys_13} is a much more recent incarnation.  It generates
automation scripts for smartphones from natural language queries.  First, it
uses natural language processing techniques to parse the queries.  Then it
applies program synthesis techniques to the parsing result to construct the
scripts.

Our \emph{vision} is to \emph{generalize} current keyword programming systems
into a new programming model where users ``program'' using a minimalistic,
universal programming language.  The programmer should be to write down thoughts
and not worry about syntax details. 

Our idea to advance this vision is inspired by the observation that natural
language is robust against noise.  The meaning of many sentences survives the
loss of words, sometimes many of them.  In other words, the sentence or one
similar can often be reconstructed given a few key words.  We call these words
``wheat'' and the rest ``chaff''.  We hypothesize that this intuitive
observation about natural language also holds for programming languages:

\noindent \textbf{Wheat and Chaff Hypothesis:} Units of code consists of 1) 
``wheat'', important lexical features that preserve meaning, and ``chaff'', and
2) the ``wheat'' is small compared to ``chaff.''


If we can \emph{distill} source code into ``wheat'', perhaps, we can gain
insights into how to \emph{expand} ``wheat'' into source code and, thus, take a
step toward realizing the new programming model we envision.  In these terms,
the programmer would write the ``wheat'' and the system would fill in the
``chaff.''

We call the phenomenon of distilling source code into a subset of lexical
features that uniquely identifies it, \emph{lexical distinguishability}.  By
studying lexical distinguishability, we are the first to provide quantitative
and qualitative evidence that the Wheat and Chaff Hypothesis holds.  The benefit
of our approach is that we establish the existence of a ``small'' subset of
words that uniquely \emph{maps} to a larger, executable piece of code; thus, we
provide evidence supporting the assumption underlying much work.  The main
limitation of our approach is that the ``wheat'' is artificial; it may not be
what a human would use in applications like code search or keyword programming.
We attempt to overcome this limitation. 

We focus our study on a diverse corpus of real-world Java projects with 100M
lines of code. The approximately 9M Java methods in the corpus form our universe
of discourse as methods capture natural, likely distinct units of source code.
Against this corpus, we compute a \emph{minimal distinguishing subset (\Minset)}
for each method.  This \Minset is the wheat of the method and the rest is chaff.
We represent each method as a bag-of-words.  We develop an algorithm to compute
their \textsc{minsets}.  A lexicon is a set of words.  Like web search queries,
\textsc{Minsets} are built from words in a lexicon.  We run our algorithms over
different lexicons, ranging from raw, unprocessed source tokens to various
abstractions of those tokens, all in a quest to find a natural, expressive and
meaningful lexicon that culminated in the discovery of a natural lexicon to use
for queries (\autoref{sec:results:natlex}).  

Our results show programs do indeed contain a great deal of chaff. Using the
most concrete lexicon, formed over raw lexemes, \textsc{minsets} compose only
$4$\% of their methods on average. This means that about $96$\% of code is
chaff.  While the ratios vary and can be large, \textsc{minsets} are always
small, containing, on average, $1.56$ words, and none exceeds $6$.  We observed
the same trend over other lexicons.  Detailed results are in
\autoref{sec:results}.  \autoref{sec:apps} also discusses existing and
preliminary applications of our work.  Our project web site
(\url{http://jarvis.cs.ucdavis.edu/code\_essence}) also contains more
information on this work, and interested readers are invited to explore it.

\vspace*{3pt}
Our main contributions follow:
\begin{itemize}
\item We define and formalize the \Minset problem for rigorously testing the
  Wheat and Chaff hypothesis (\autoref{sec:threshing:winnowing});
\item We prove that \Minset is \emph{NP-hard} and provide a greedy
  algorithm to solve it (\autoref{sec:threshing:minset}); 
\item We validate our central hypothesis --- \emph{source code
  contains much chaff} --- against a large (100M LOC), diverse corpus
  of real-world Java programs (\autoref{sec:results}); and
\item We design and compare various lexicons to find one that is natural,
expressive, and understandable (\autoref{sec:results:natlex}).
\end{itemize}

The rest of this paper is organized as follows. In \autoref{sec:threshing} we
define lexical distinguishability of source code and explain how we study it.
\autoref{sec:methodology} describes our Java corpus, and implementations of the
feature extractor and the \Minset algorithm. \autoref{sec:results} presents our
detailed quantitative and qualitative results. \autoref{sec:discussion} analyzes
our results and their implications. \autoref{sec:relwork} places our work into
the context of related work, and \autoref{sec:concfw} concludes.

%% file: threshing_and_winnowing.tex
\section{Problem Formulation}
\label{sec:threshing}

In this section, we describe how we determine if a piece of code is lexically
distinguishable.  We explain our representation of code.  We also introduce
several definitions including \emph{distinguishing subset}, \Minset, and the
\Minset problem.  Finally, we present and discuss our \Minset algorithm. 

\subsection{Bag-of-Words Model}
\label{sec:threshing:threshing}

The first step in our formulation is to define the unit of code.  One could
choose units like individual statements, blocks, functions, or classes.  In this
study, we view functions as the units of code.  This granularity seems adequate.
Functions are natural, likely distinct, pieces of code and functionality.
Functions are also reusable building blocks of more complex components.  

We represent a unit of code, function, as a set of lexical features or
bag-of-words.  We disregard syntactic structure, order, and multiplicity.
First, we parse each function to get its set of lexemes.  A lexeme is a
delimited string of characters in code, where space and punctuation are typical
delimiters; it is an atomic syntactic unit in a programming
language.\footnote{Linguistics defines a lexeme differently.  A lexeme is the
\emph{set} of forms a single word can take.  For example, `run', `runs',
`running' are all forms of the same lexeme identified by the word `run'.} Then,
we map each lexeme to a \emph{word}.

\boldpara{What is a ``word''?} A \emph{word} is a lexeme, or some abstract or
refined form of it.  A \emph{lexicon} is a set of words.  For example, a
natural, basic lexicon is the set of raw lexemes.  Using this lexicon, the
mapping of lexemes-to-words would be simple.  Each lexeme would map to itself.
The bag-of-words for each function would be its set of lexemes.  

\subsection{Lexicons}
\label{sec:threshing:lexicons}
What is a word depends on the choice of the lexicon.
The freedom to define the lexicon allows us to sharpen, blur, or even disregard
certain lexical features.

New lexicons can be formed by abstraction over lexemes.  In natural languages,
for example, the words in a sentence can be replaced by their part of speech,
like \Noun, \Verb, or \Adj, to highlight phrase structure.  Similarly, code
parsers tag each lexeme with one of a set of token types.  For example, the
javac lexer defines $101$ token types, for example, \ID and
\INTLIT~\cite{openjdk_12}.  This set of token types is another natural but
clearly more abstract lexicon.  Using this lexicon, we would map each lexeme to
its token type.  For example, ``\lstinline{3.14}'' would map to \INTLIT.  A word
would be one of the $101$ token types.  The bag-of-words for each function would
be a subset of these $101$ token types.

New lexicons can also be defined by filtering specific lexemes.  For example, we
can define a lexicon consisting of all lexemes except separators, like
``\lstinline{(}'' and ``\lstinline{)}''.  Using this lexicon, we would map each
lexeme to itself except separators.  Separator lexemes would map to nothing.
The bag-of-words for each function would be it's set of lexemes minus the
separator lexemes.

\boldpara{Homonyms}
Functions may contain, to adapt a word from linguistics, \emph{homonyms}:
identical lexemes with distinct effects on behavior.  For example, in Java, the
lexeme ''\lstinline{get}'' could be a method call of
``\lstinline{java.util.Map.get()}'' or ``\lstinline{java.util.List.get()}''.  In
Java, we can fully qualify homonyms to distinguish them.

\boldpara{Synonyms} 
We can preserve lexical differences that we suspect capture differences in the
behavior of a method by ensuring that different lexemes map to distinct words.
We can also blur lexical differences by abstracting distinct lexemes we suspect
have the same effect on behavior, \ie \emph{synonyms}, to the same word.  For
example, variable identifiers can be replaced with their type under a language's
type system.  In the top method shown in \autoref{lst:sample}, the parameter
``\lstinline{array}'' could just as well have been named ``\lstinline{values}''.

In general, a lexicon that is fine-grained and concrete may exaggerate
unimportant differences between functions, while one that is coarse and abstract
may blur important differences.  
Varying the lexicon allows us to explore programming language-specific
information.  The lexicon consisting of all lexemes probably includes many
elements that distinguish but probably have little to do with the behavior of
functions, \ie, delimiters and string literals like \lstinline{"Joe"}.  We can
filter those lexemes.  We can also filter other lexemes, like the type
annotation ``\lstinline{int}'' in ``\lstinline{int cars = 0;}'', to explore how
distinguishing they are.

\subsection{Illustration of the Bag-of-Words Model}


\begin{lstlisting}[language=Java, caption={This listing shows two Java methods.
Both implement the BubbleSort algorithm.  (top) Sorts an array of integers. 
(bottom) Sorts an array of strings.}, label={lst:sample}, float=t]
/**
 * Standard Bubble Sort algorithm.
 * @param array The array to sort.
 */
private static void bubbleSort(int array[]) {
	int length = array.length;
	for (int i = 0; i < length; i++) {
		for (int j = 1; j > length - i; j++) {
			if (array[j-1] > array[j]) {
				int temp = array[j-1];
				array[j-1] = array[j];
				array[j] = temp;
			}
		}
	}
}

static void bubblesort(String[] values) {
	// no Java sort, so ugly bubble sort
	for (int i=0; i<=values.length-2; i++) { // stop sort early to save time!
		for (int j=values.length-2; j>=i; j--) {
			// check that the jth value is smaller than j+1th,
			// else swap
			if (0 < values[j].compareTo(values[j+1])) {
				// swap
				String temp = values[j];
				values[j] = values[j+1];
				values[j+1] = temp;
			}
		}
	}
}

\end{lstlisting}

\autoref{lst:sample} shows two Java methods found in real-world projects, Apache
Log4j and JMRI (A Java Model Railroad Interface), respectively.  The first
method sorts an array of integers.  The second one sorts an array of Strings.
They both sort using the Bubble Sort algorithm. 

\begin{figure}[t]
  \centering
	\includegraphics[width=0.48\textwidth]{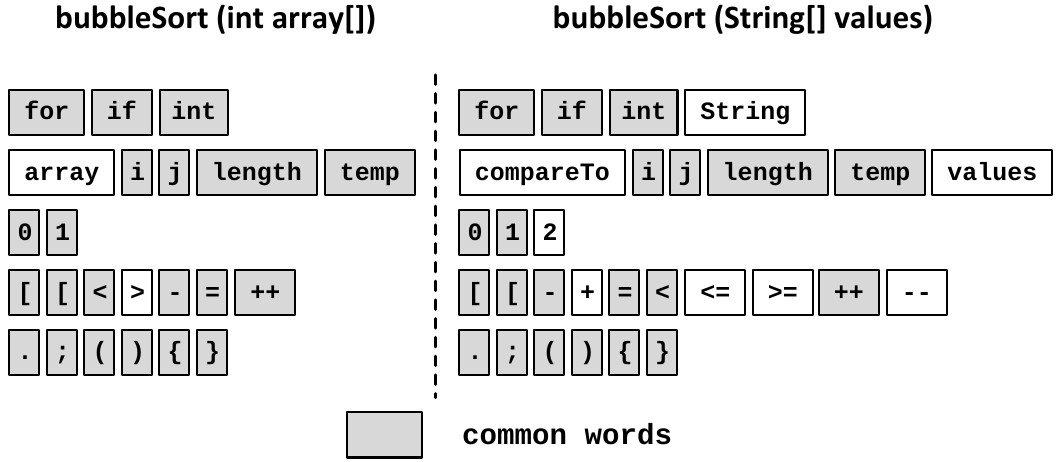}  \\
\caption{(left) This is the simplified bag-of-words representation of the method
that sorts an array of integers using Bubble Sort.  (right) This is the
simplified bag-of-words representation of the method that sorts an array of
Strings using Bubble Sort. Note: The words in common are shaded.} 
  \label{fig:samplebags}
\end{figure}

\autoref{fig:samplebags} shows their simplified representation as a
bag-of-words.  For this example, we have defined the lexicon to be all lexemes.
Thus, the words are simply the raw lexemes.  To help visualize the similarity
between these two methods, we have shaded the words in common; there have $21$
words in common.  This should not be surprising.  Both methods implement the
same functionality.  The main difference is that they operate over elements of
different types, ``\lstinline{int}'' and ``\lstinline{String}''.  By shading the
common words, we also highlight the differences between these two methods.  For
example, the second method uses the ``\lstinline{--}'' (decrement) operator in
the second loop to iterate backwards.

\subsection{Distinguishable Code}
\label{sec:threshing:winnowing}

We simplified the representation of a function by mapping its source code to a
set of lexical features, \emph{bag-of-words}.  Finding what distinguishes a
function lexically is thus reduced to finding a unique subset of code features
or words.  This unique subset distinguishes each function from all other
functions (when each functions is represented as a bag-of-words).  We call any
such subset a \emph{distinguishing subset}, and define it precisely in
Definition \autoref{defn:ds}.  A function may not have a distinguishing subset. 
We call those that do, \emph{distinguishable} (Definition \autoref{defn:dc}).

\begin{defn}
\label{defn:ds}
Given a finite set $S$, and a finite collection of finite sets $\mathcal{C}$,
$S^*$ is a \emph{distinguishing subset} of $S$ if and only if
\begin{tabular}{l>{$}l<{$}l}
(P1) & S^* \subseteq S   & $S^*$ is a subset of $S$ \\
(P2) & \forall C \in \mathcal{C},\ S^* \not\subseteq C
	 & $S^*$ is \emph{only} a subset of $S$ \\
\end{tabular} \\
\end{defn}

\begin{defn}
\label{defn:dc}
A unit of code is \emph{lexically distinguishable} if it has a distinguishing
subset.
\end{defn}

\boldpara{The \Minset problem} 
A unit of code may have more than one distinguishing subset.  To determine if it
is distinguishable, we simply need to find one. 
We focus on finding a \emph{minimum distinguishing subset}
(\Minset). 
We call this The \Minset Problem (Definition \autoref{defn:mp}).
It is the \emph{core} computational problem that we study.

\begin{defn}[The \Minset Problem]
\label{defn:mp}
Given a finite set $S$, and a finite collection of finite sets $\mathcal{C}$,
find a \emph{minimum} distinguishing subset (\emph{minset}) $S^*$ of $S$.
\end{defn}

\begin{thm}
\Minset is NP-hard.
\end{thm}

\begin{proof}
We reduce \HS to $\Minset$.  
\end{proof}

A \Minset identifies a piece of code.  It consists of lexically distinguishing
features.  Some features may crucially differentiate its behavior from similar
functions.  Some may not.  A \Minset, however, is not itself executable.  It
depends on its surrounding context to execute and provide functionality.  

In the keyword-query sense, a \Minset is the smallest query that will uniquely
identify and recall a piece of code.  It may not be what humans would actually
attempt to use.  That is a separate challenge.  In this study, we focus on
finding and studying minsets.

\subsection{The \textsc{\large Minset} Algorithm}
\label{sec:threshing:minset}

\begin{algorithm}[t]

\caption{Given the universe $U$, the finite set $S$, and the finite set of
finite sets $\mathcal{C}$, $\Minset$ has type $2^U \times 2^{2^U} \rightarrow
2^U \times 2^{2^U}$ and its application $\Minset(S,\mathcal{C})$ computes 1)
$S^* \subset S$, a subset that distinguishes $S$ from sets in $\mathcal{C}$, and
2) $\mathcal{C}'$, a ``remainder'', \ie a subset of $\mathcal{C}$ whose sets
contain $S$ and therefore from which $S$ could not be distinguished; when
$\mathcal{C}' = \emptyset$, $S^*$ distinguishes $S$ from all the sets in
$\mathcal{C}$'; when $\mathcal{C}' = \mathcal{C}$, $S^* = \emptyset$.}

\label{alg:minset}
\begin{algorithmic}[1]
  \REQUIRE $S$, the set to minimize.
  \REQUIRE $\mathcal{C}$, the collection of sets against which $S$ is minimized.\!
  \STATE $\mathcal{C}_e = \{C \mid C \in \mathcal{C} \wedge e \in C \}$ are
  those sets in $\mathcal{C}$ that contain $e$.
  \STATE $S^* = \emptyset$
  \WHILE{$S \ne \emptyset \wedge \mathcal{C} \ne \emptyset$}
    \LCOMMENT{Greedily pick an element that most differentiates $S$.}
    \STATE $e := \textsc{choose}(
    \{x \in S \mid |\mathcal{C}_x| \leq |\mathcal{C}_y|, \forall y \in S\}
    )$
    \STATE \textbf{if } $\mathcal{C}_e = \emptyset 
        \vee \mathcal{C}_e = \mathcal{C}$ \textbf{ break}
    \STATE $S^* := S^* \cup \{e\}$
    \STATE $S := S \setminus \{e\}$
    \STATE $\mathcal{C} := \mathcal{C}_e$
  \ENDWHILE
  \RETURN $S^*, \mathcal{C}$
\end{algorithmic}
\end{algorithm}

Since the \Minset problem is NP-hard, we present \algref{alg:minset}, a greedy
(approximation) algorithm that finds the locally minimal distinguishing subset
of a set $S$.  Given inputs $S$, the target set to be minimized, and
$\mathcal{C}$, a collection of sets against which $S$ is minimized, the \Minset
algorithm computes $S^*$, and $\mathcal{C}'$.  $C'$ is the subset of
$\mathcal{C}$ whose sets contain $S$ so $\mathcal{C} \setminus \mathcal{C}'$
contains those sets in $\mathcal{C}$ that do not contain $S$.  When
$\mathcal{C}' = \emptyset$, $S^*$ is a subset of $S$ that distinguishes $S$ from
all sets in $\mathcal{C}$.  The core of the algorithm is Line 4.  Equality is
needed in the cardinality test for cases like $S = \{a,b\}, \mathcal{C} =
\{\{a,x\},\{a,y\},\{b,x\},\{b,y\}\}$, where all the elements in $S$
differentiate $S$ from the same number of sets in $\mathcal{C}$.  Equality also
means that $C_x$ can be empty, as for   $S = \{a\}$ and $\mathcal{C} =
\{\{x\},\{y\}\}$, since $|C_a| \leq |C_a| = 0$, and $C_x$ can also be
$\mathcal{C}$ again, when $S \subseteq C, \forall C \in \mathcal{C}$, as in $S =
\{a\}$ and $\mathcal{C} = \{\{a\},\{a,b\},\{a,b,c\}\}$.

\begin{figure}[t]
  \centering
	\includegraphics[width=0.48\textwidth]{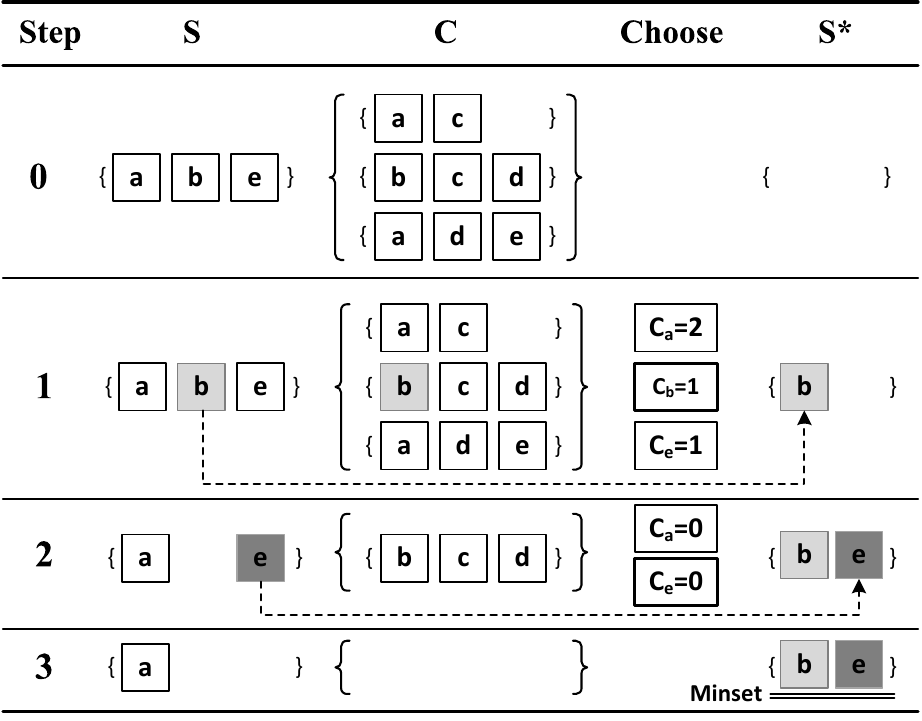}  \\
\caption{The execution of \algref{alg:minset} illustrated on the following 
problem instance: $\Minset(\{a,b,e\},\{\{a,c\},\{b,c,d\},\{a,d,e\}\})$.}
\label{fig:minset:ex}
\end{figure}

\autoref{fig:minset:ex} shows a sample run of the algorithm.  It ends in two
iterations.  It finds a \Minset of $\{a,b,e\}$ with respect to the collection
$\mathcal{C} = \{\{a,c\},\{b,c,d\},\{a,d,e\}\}$.  The \Minset is $\{b,e\}$.
None of the sets in $\mathcal{C}$ contain this \Minset.

\begin{thm}
Consider $\Minset(S,C) = S^*,\mathcal{C}'$. The $S^*$ that \algref{alg:minset}
computes distinguishes $S$ from a subset of $\mathcal{C}$;  when $\mathcal{C}' =
\emptyset$, $S^*$ is a locally minimal distinguishing subset of $S$.
\end{thm}

\begin{proof}
By induction on $S^*$.
\end{proof}

The worst case time complexity of $\Minset(S,\mathcal{C})$ is
$O({|S|}^2|\mathcal{C}|)$.  First, there are $|S|$ iterations and, in each call,
for each element $x \in S$, we need to, 1) compute $\mathcal{C}_x$, each at a
cost of $|\mathcal{C}|$, for a total cost of $O(|S||\mathcal{C}|)$, then 2) then
find the minimum $|\mathcal{C}_x|$ at a cost of $O(|S|)$.  Of course, $S$ and
$\mathcal{C}$ are smaller in each iteration, but we ignore this and
over-approximate.  Thus, we have $O(|S|(|S||\mathcal{C}| + |S|)) =
O({|S|}^2|\mathcal{C}|)$.

As mentioned earlier, modeling functions as sets discards differences in methods
due to multiplicity.  We have also developed a multiset version of the \Minset
algorithm, which we omit due to lack of space.

%% file: methodology.tex
\section{Setup and Implementation}
\label{sec:methodology}

We selected a very popular, modern programming language, Java, and collected a
large ($100$M lines of code), diverse corpus of real-world projects.  Ignoring
scaffolding and very simple methods, which we define as those containing fewer
than 50 tokens, there are $1,870,905$ distinct methods in our corpus.  We
selected a simple random sample of $10,000$ methods\footnote{Given the
population size, this sample size gives us a confidence level of $95$\%, and a
margin of error of $\pm1$\% in our measures.}.  Our software and data is
available
online\footnote{\label{foot:dl}\scriptsize{\url{https://bitbucket.org/martinvelez/code\_essence\_dev/downloads}}.}.

\subsection{Code Corpus}
\label{sec:method:corpus}

\begin{table}[t]
\caption{Corpus summary.}
\label{tab:corpus}
\centering
\begin{tabular}{lrrr}
\hline
\toprule
Repository & Projects   & Files & Lines of Code                \\
\midrule
Apache              &  103          & 101,480 & 10,891,228          \\
Eclipse             &  102          & 287,669 & 32,770,246          \\
Github              &  170          & 133,793 & 13,752,295          \\
Sourceforge         &  533          & 373,556 & 42,434,029          \\
\midrule
\textbf{Total}      &  908          & 896,498 & 99,847,798          \\
\bottomrule
\end{tabular}
\end{table}

We downloaded almost one thousand of the most popular projects from four
widely-used open source code repositories: Apache, Eclipse, Github, and
Sourceforge.

\boldpara{Curation} 
Since some projects in our corpus are hosted in multiple code repositories, we
removed all but the most recent copy of each project.  Also, since many project
folders contained earlier or alternative versions of the same project, and even
other projects, where we could, we identified the main project and kept only its
most current version.  \autoref{tab:corpus} summarizes our curated corpus.
After curation, clones still existed in the corpus, for example, within
projects.  A search program we wrote helps us find clones.  When we compute
minsets, we assume no clones remain. Our results in \autoref{sec:results:size}
give us confidence that this is the case.  

\boldpara{Filtering Scaffolding Methods} Java, in particular, requires that a
programmer write many short scaffolding methods, for example, getters and
setters.  Many languages, like Ruby and Python, eliminate the need for such
scaffolding code.  After manual inspection, we found that such methods usually
contain less than $50$ tokens, or about $5$ lines of code.  This is consistent
with other research~\cite{li2004cpminer, basit2007tokenclonedetection} that also
ignores shorter methods.  At this size, we also filter methods with very simple
functionality.  After filtering, $905$ out of $908$ projects are still
represented.  \autoref{tab:methods} shows the method counts.  

\begin{table}[t]
\caption{Method counts.}
\label{tab:methods}
\centering
\begin{tabular}{l r}
\hline
\toprule
Methods & Count \\
\midrule
Total (in corpus) 							& $8,918,575$ \\
Unique 													& $8,135,663$ \\
Unique ($50$ or more tokens) 		& $1,870,905$ \\
Unique ($50$ to $562$ tokens)		& $1,801,370$ \\
\bottomrule
\end{tabular}
\end{table}

\subsection{The Feature Extractor}

We developed a tool, which we call JavaFE, that processes all the functions in
our corpus.  JavaFE leverages the Eclipse JDT parser which parses Java code and
builds the syntax
tree\footnote{\scriptsize{\url{http://www.eclipse.org/jdt/}.}}. JavaFE can take
as input \texttt{.java}, \texttt{.class}, and \texttt{.jar} files.  Projects can
contain these and other types of files.  The tool builds a list of tokens for
each method.  It collects the lexeme of each token and additional information as
it traverses the syntax tree.  

To address the \emph{homonym} problem, JavaFE collects the fully qualified
method name (FQMN) for method name lexemes, and the fully qualified type name
(FQTN) for variable identifiers and type identifiers.  Collecting this
information allows us later to classify methods and types based whether they are
part of the Java SDK library or if they are local to specific projects.  When
projects are missing dependencies, resolving names to either FQMN or FQTN may
not be possible.  In our corpus, we encountered this problem with $0.03$\% of
the tokens.  JavaFE can also collect more abstract information like lexer token
types as defined in the \texttt{javac} implementation of OpenJDK, an open-source
Java platform~\cite{openjdk_12}.  

\subsection{The \Minset Algorithm Implementation}

All the information collected by JavaFE is stored in a PostgreSQL database.  We
developed a Ruby program that runs the \Minset algorithm for each method and
stores the result in the same database.  If a method does not have a minset, it
stores a list of its strict supersets, and a list of methods that are duplicates
when represented as a bag-of-words.

%% file: results.tex
\section{Results and Analysis}
\label{sec:results}

\begin{table}[t]
\footnotesize
\caption{Lexicons.}
\label{tab:lexicons}
\begin{tabular}{l l r}
\toprule
Name  & Description & Size (words) \\
\midrule
\Lex & All (raw) lexemes & 5,611,561 \\
\LTT & All lexer token types & $101$ \\
\MinOne & Fully qualified standard library method names & 55,543\\
 & and basic operators & 55,543 \\ 
\MinTwo & \MinOne plus control keywords & 55,556 \\ 
\MinThree & \MinTwo plus fully qualified public type names & 91,816 \\ 
\MinFour & \MinThree plus additional keyword and token types & 91,829 \\
\bottomrule
\end{tabular}
\end{table}

\begin{figure*}[t]
  \centering
	\includegraphics[width=0.92\textwidth]{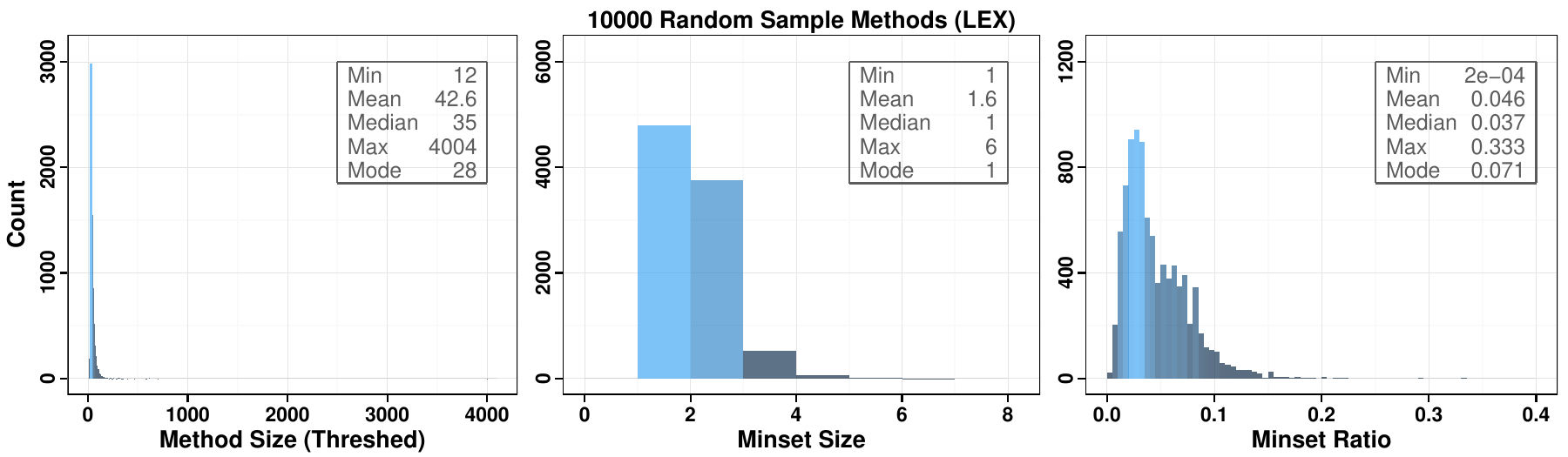} \\ 
\caption{The histogram of minset sizes tells us that minsets are small.
Comparing minset sizes with method sizes shows that minsets are also relatively
small.  The minset ratio histogram confirms this.}
  \label{fig:min50-lex-random}
\end{figure*}

\begin{figure}[t]
  \centering
  \includegraphics[width=0.48\textwidth]{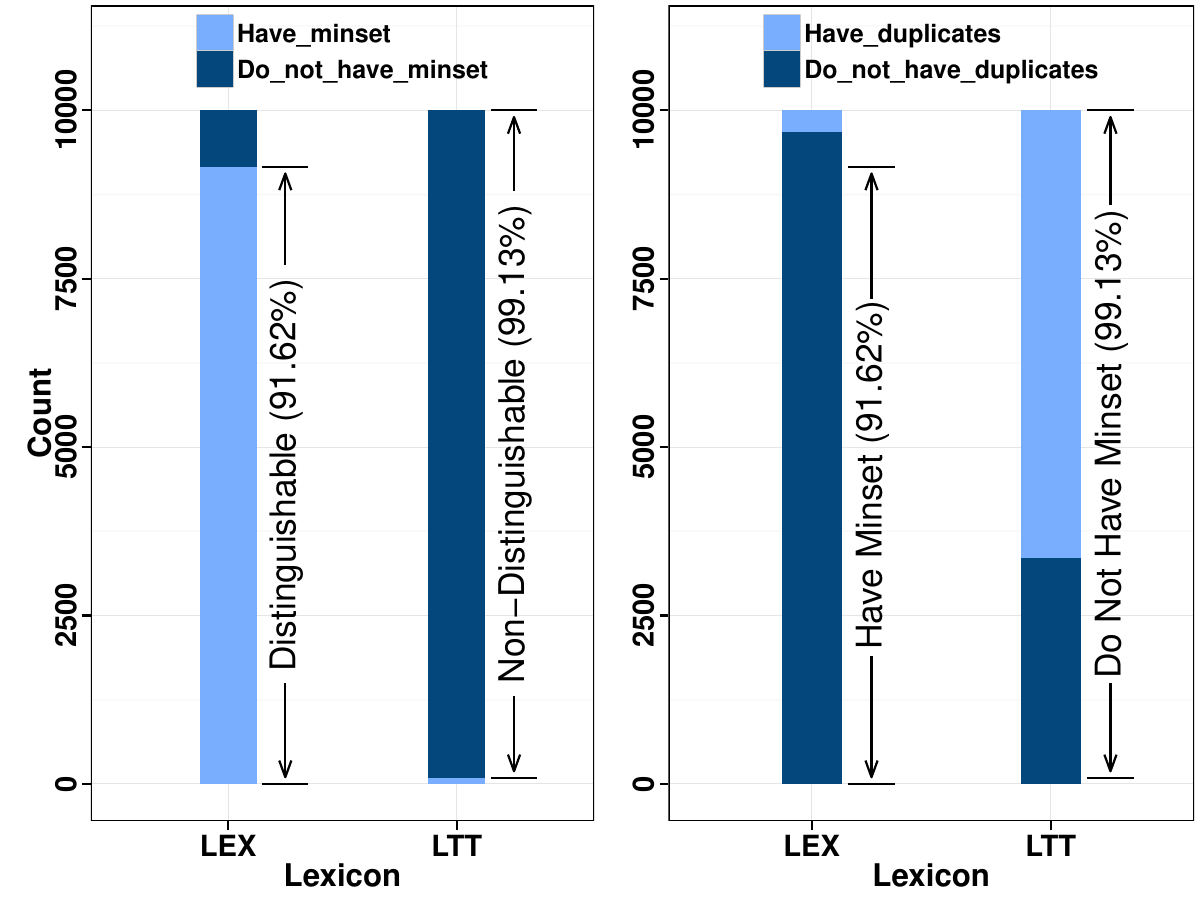}  \\
	\caption{Random Sample of $10,000$ Methods: (left) \emph{Proportion of Methods
with Minsets:} There is a stark difference in that proportion between \Lex and
\LTT.  (right) \emph{Proportion of Methods with Duplicates:} \Lex induces very
few duplicates compared to \LTT.  \LTT maps almost three quarters of the methods
to the same set as another.  It is too coarse, and does not model methods well.}
  \label{fig:yield10000}
\end{figure}

We provide quantitative and qualitative results for the following questions:
\begin{enumerate}
\item How many units of code are lexically distinguishable?
\item How much of code is needed to distinguish it?
\item To what extent do minsets also capture code behavior and behavior differences?
\item What is a natural, minimal lexicon?
\end{enumerate}

\boldpara{Measures\footnotemark[2]} 
We define the \emph{yield} of a lexicon to be the percentage of distinguishable
methods in our corpus.  The second question can be addressed in terms of
absolute \emph{minset size}, or in terms of \emph{minset ratio}, minset size to
threshed method size.  While minset sizes and minset ratios will almost
undoubtedly vary across functions, we hypothesize that the \emph{mean} minset
size and the \emph{mean} minset ratio are small.

\boldpara{Lexicons}  We provide results over $6$ different lexicons, listed in
\autoref{tab:lexicons}.  \Lex and \LTT are the lexicons we discussed in
\autoref{sec:threshing:lexicons}. \MinOne - \MinFour are lexicons we explore in
the search for a natural, minimal lexicon.

\boldpara{Summary} 
Code is lexically distinguishable.  Perhaps just as importantly, only $1.56$
words, on average, or just $4$\%, are needed to distinguish a unit of code from
all others in the corpus over \Lex.  The problem with minsets over \Lex is that
they do no capture behavior and behavior differences well.  Over \MinFour, on
the other hand, minsets are still small but reveal much more about the behavior
of the code because we intentionally blurred lexical differences which we
suspect do not distinguish behavior.  We elaborate on this point in
\autoref{sec:results:natlex} and \autoref{sec:results:semantics}. 

All of our data and data processing code can be downloaded from
Bitbucket.\footnotemark[3]

\begin{table*}[t]
\footnotesize
\caption{Types of lexemes (or words) in the minsets we computed over the lexicon
\Lex.}
\label{tab:lexemetypes}
\centering
\resizebox{17.5cm}{!}{
\begin{tabular}{ l r l }
\toprule
Grain Type														& Count	& Examples \\
\midrule
Variable Identifier (of Public Type) 	& 3235 	& abilityType (java.lang.StringBuffer), defaultValue (int), lostCandidate (boolean), twinsItem (java.util.List) \\
String and Character Literal 					& 3202 	& `\textbackslash{}u203F', `\&', "192.168.1.36", "audit.pdf", "Error: 3", "Joda", "Record Found", "secret4" \\
Method Call (Local)										& 2942 	& classNameForCode, getInstanceProperty, isUserDefaultAdmin, makeDir, shouldAutoComplete \\
Variable Identifier (of Local Type)		& 1574 	& arcTgt, component, iVRPlayPropertiesTab, nestedException, this\_TemplateCS\_1, wordFSA \\
Type Identifier (a Local Type)				& 1413	& ErrorApplication, IWorkspaceRoot, Literals, NNSingleElectron, PickObject, TrainingComparator \\
Method Call (a Public Method)					& 508		& currentTimeMillis (java.lang.System.currentTimeMillis()), replace (java.lang.String.replace(char,char)) \\
Number Literal (integer, float, etc.)	& 310 	& 0, 1, 3, 150, 2010, 0xD0, 0x017E, 0x7bcdef42, 255.0f, 0x1000000000041L, 46.666667 \\
Type Identifier (a Public Type)				& 265		& int, ArrayList, Collection, IllegalArgumentException, PropertyChangeSupport, SimpleDateFormat \\
Operator															& 260 	& \textasciicircum{}=, <, \lstinline+<<=+, <=, =, ==, >, >=, \lstinline+>>+, \lstinline+>>=+, \lstinline+>>>=+, |, |=, ||, -, -=, --, !, !=, ?, /, /=, @, *, \&, \&\&, +, +=, ++ \\
Keyword	(Except Types)								& 196 	& break, catch, do, else, extends, final, finally, for, instanceof, new, return, super, synchronized, this, try, while \\
Separator															& 148 	&  <, >, ", ", ., ] \\
Reserved Words (Literals) 						&	104		&  false, null, true \\
Other																	&	112		& COLUMNNAME\_PostingType, E, ec2, element, ModelType, org, T, TC \\
\bottomrule
\end{tabular}
}
\end{table*}

\subsection{Lexical Distinguishability of Source Code}
\label{sec:results:size}

The question ``How much of a piece of code is needed to distinguish it from
others?'' can be answered in  two ways: in terms of \emph{minset size} and
\emph{minset ratio}.  We report both.  

There are two natural views we can take of code: the raw sequence of lexemes the
programmer sees when writing and reading code, and the abstract sequence of
tokens the compiler sees in parsing code.  We want to explore those two views,
and capture each one as a lexicon, a set of words. \Lex is the set of all
lexemes found in code ($5,611,561$ words).  \LTT is the set of lexer token types
defined by the compiler ($101$ words).  Each word in \LTT is an abstraction of a
lexeme, like \lstinline+3+ into \lstinline+INTLIT+. 

\boldpara{\Lex} \Lex is the primordial lexicon;  all others are abstractions of its words.
Unfortunately, it is noisy:  it is sensitive to any syntactic differences,
including typos or use of synonyms, so it tends to overstate the number of
minsets and understate their sizes; spurious homonyms can have the opposite
effect, but are unlikely in Java when one can employ fully qualified names.
\LTT is the minimal lexicon a parser needs to determine whether or not a string
is in a language.  We computed minsets of all the methods in our random sample
of $10,000$ using each lexicon, and display a summary of our results in
\autoref{fig:min50-lex-random} and \autoref{fig:yield10000}.

Using \Lex, a tiny proportion of code is needed to distinguish it.  The minset
of a method, on average, contains $4.57$\% of the unique lexemes in a method
which means that methods in Java contain a significant amount of chaff,
$95.43$\% on average.  More surprisingly, the number of lexemes in a minset is
also just plain small.  The mean minset size is $1.55$.  The minset sizes also
do not vary much.  In $85.62$\% of the methods, one or two unique lexemes
suffices to distinguish the code from all others.  The largest minset consists
of only $6$ lexemes.  Minset ratios also do not vary much.  $75$\% of all
methods have a minset ratio of $6.35$\% or smaller.  While the ratios are
sometimes large, the absolute sizes never are.  The method with the largest
minset ratio, $33.3$\%, for example, consists of $18$ unique lexemes but has a
minset size of $6$.  The method with the second largest minset ratio, $29.41$\%,
another example, consists of $17$ unique lexemes and has a minset size of $5$.

\boldpara{Minset Sizes of Large Methods} 
Minsets are surprisingly small; especially surprising is that the maximum size
is small.  One reason might be the compression inherent to representing
functions as sets.  We address this later when we experiment with multisets.  To
test the robustness of our results, we also focused our investigation on larger
methods because they may encode more behavior and therefore have more
information.  Hence, they may have larger minsets.  Selected uniformly at
random, our sample set does not include many of the largest methods: the largest
method in our random sample has $2025$ lines of code while the largest one in
our corpus contains $4,606$ lines of code. To answer this question about minset
properties conditioned on large methods, we selected the $1,000$ largest
methods, by lines of source code, and computed their minsets.  The mean and
maximum minset sizes of the largest methods are slightly lower but similar to
the previous sample, $1.12$ and $4$, respectively.  This shows that minsets are
small and potentially effective indices of unique information even for
abnormally large methods.

\boldpara{\LTT} 
Using \LTT, the proportion of words needed to distinguish code is larger but
still small.  The minset of a method, on average, contains $18.45$\% of the
unique token types in a method.  We observe again that sometimes minset ratios
can be large but the absolute minsets sizes never are.  It is not surprising
that the minset ratio is larger.  Information is lost in mapping millions of
distinct lexemes to only $101$ distinct lexer token types.  Information is also
lost as method sizes decrease from $42.7$ using \Lex to $18.2$ using \LTT.

These results show that few words are needed to distinguish code, in relative
and absolute terms.  Given that we preserve a lot of information with \Lex, we
claim that the mean minset size, and mean minset ratios we found are approximate
lower bounds.  In essence, we define a lexicon spectrum where \Lex is one of the
poles, and \LTT is a more abstract point on the lexicon spectrum.

\boldpara{Yield} 
The \emph{yield} of a lexicon is the percentage of distinguishable methods.  Our
exploration shows that the yield decreases as the lexicon becomes coarser,
measured roughly by the number of words in the lexicon.  Our coarsest lexicon,
\LTT, blurs lexical differences too much.  Over \LTT, only $87$ out of $10,000$,
$0.87$\%, methods have a minset.  This is in great part due to the fact that
\LTT induces many duplicates.  Over \LTT, $6,640$ out of $10,000$ are modeled to
the same bag-of-words as another method in the corpus.  Recall that all of these
methods are unique at the source code level.  In contrast, \Lex appears to
preserve sufficient lexical differences so that $9,087$ out of $10,000$ methods
have a minset.

\subsection{Minsets over \Lex}
Since there are thousands of minsets, we take a broad view of minsets.  For all
minsets, we partitioned lexemes by type, leveraging information collected
JavaFE; the types we defined are similar to lexer token types but broader in
some cases and narrower in others.  We provide a list of the lexeme types we
defined, along with the counts of lexemes belonging to that type in
\autoref{tab:lexemetypes}\footnote{A caveat: \algref{alg:minset} at line $4$
picks arbitrarily between two equally rare words. Thus, these counts could
differ.}.

Public type variable identifiers, and string and character literals dominate
minsets. String literals are constant string values like \lstinline{"Joda"}.
The strings can represent error or information messages, IP addresses, names,
pretty much anything.  Perhaps this is why are at the top of the list: they can
be unique or very rare.  We divide certain classes of words depending if they
are public or local --- method invocations, type identifiers, and variable
names.  Public words are more standard and common whereas local words are more
specialized and rare.  Not surprisingly then, we observe that standard language
features, like keywords and operators, and public types and methods are less
common in minsets.  The only exception is variable identifiers of types local to
their respective project.  Their distinctiveness is due in part to synonyms and
homonyms.  A programmer has great freedom in creating them.  For example,
\lstinline{dir} appears $8017$ times, as a variable name in methods, while
\lstinline{directory} appears only $2774$ times.  Another reason is that
variable identifiers are more prevalent than other type of identifiers, like
types and method calls.

\subsection{What is a Natural, Minimal Lexicon?}
\label{sec:results:natlex}

\begin{figure}[t]
  \centering
  \includegraphics[width=0.48\textwidth]{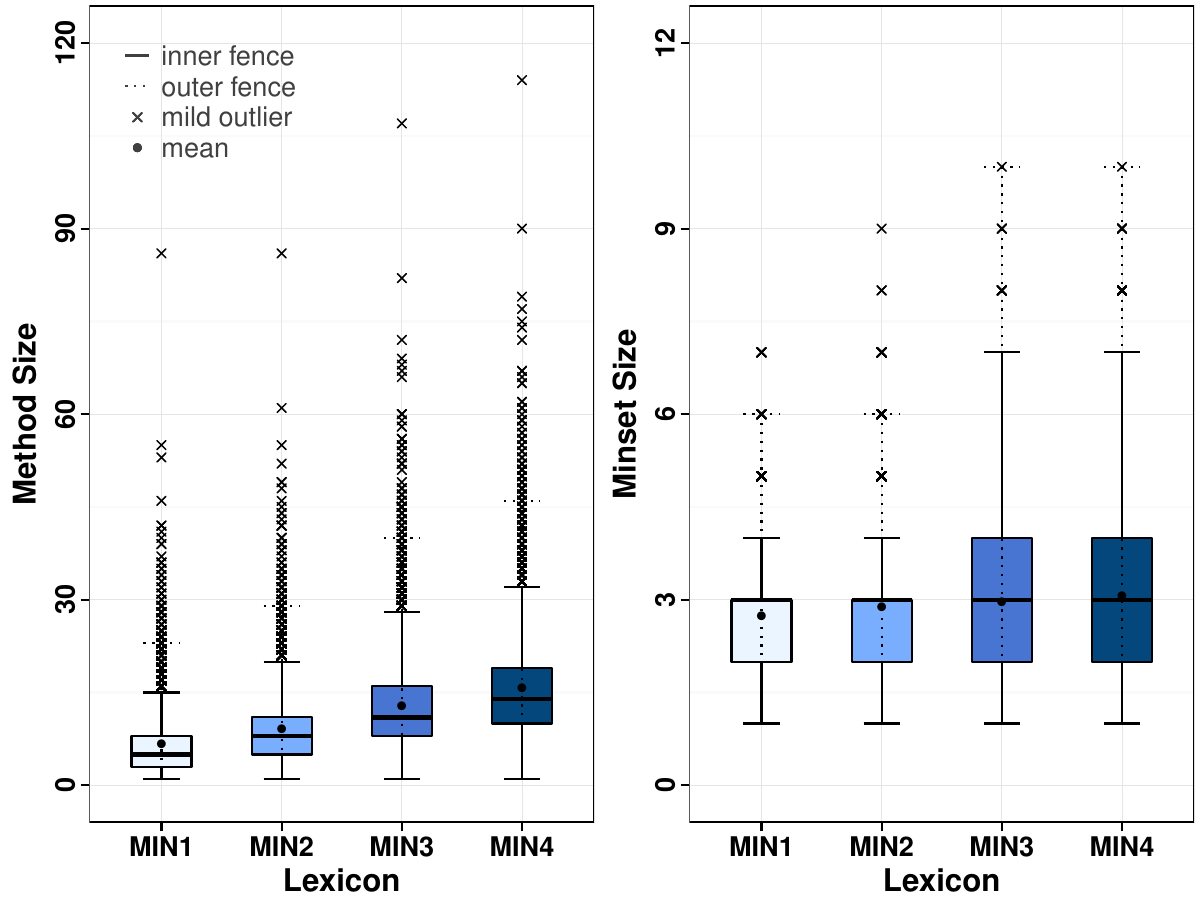}  \\
	\caption{(left) As the lexicon grows from \MinOne to \MinFour, the average
size of the threshed methods also grows.  (right) As the lexicon grows, the
average minset size hardly changes.  At least three quarters of the methods have
a minset smaller than $4$.  Even as the lexicon grows, the maximum minset size
is never more than $10$.}
  \label{fig:min50-min14}
\end{figure}

\begin{figure}[t]
  \centering
  \includegraphics[width=0.48\textwidth]{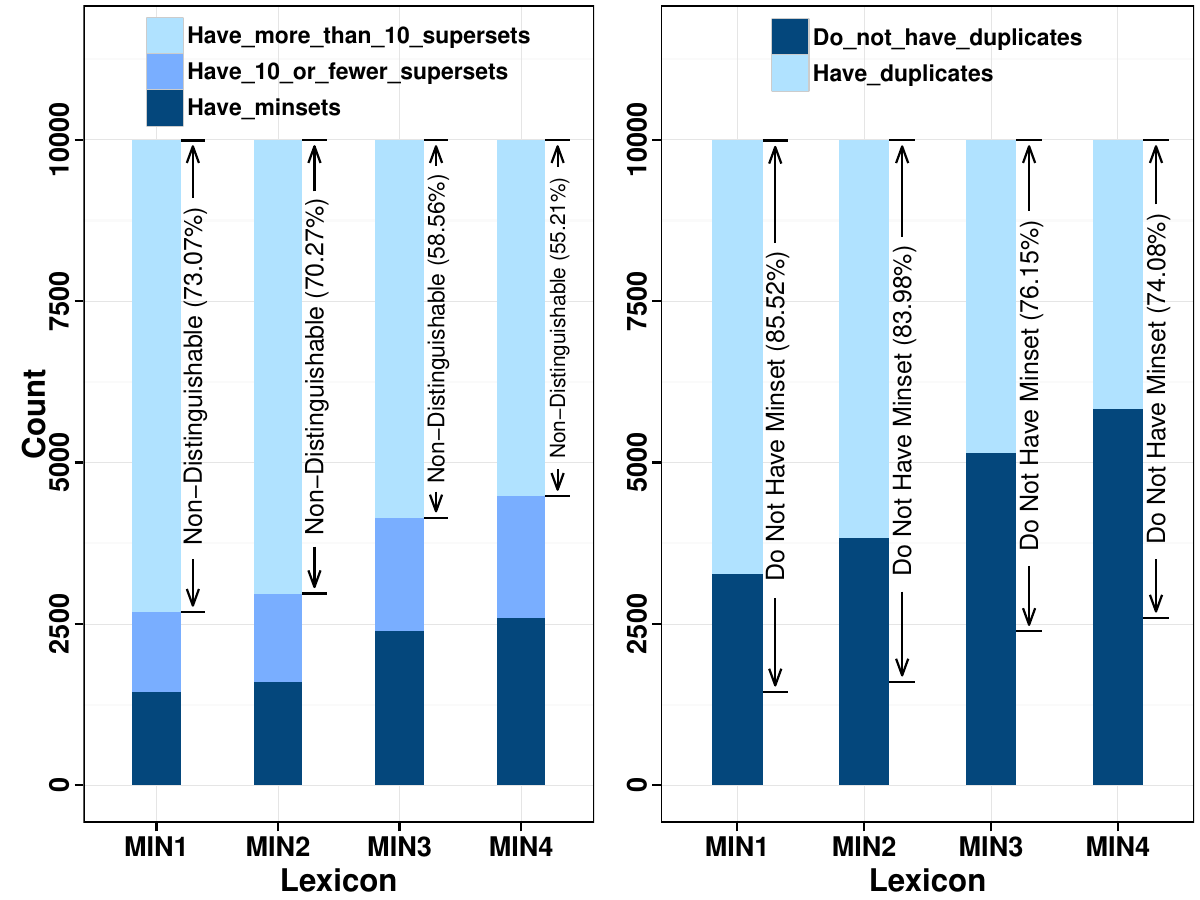}  \\
	\caption{(left) \emph{Yield:} The yield clearly improves with each change.  At
\MinFour, the yield is $44.79$\%.  (right) \emph{Proportion of Methods With
Duplicates:} Using this proportion as a rough gauge of threshing precision,
there is a substantial improvement in threshing precision with each lexicon ---
fewer methods have duplicates.  \MinFour pushes that precision past $50$\%.}

  \label{fig:min50-min14-yield}
\end{figure}

We have shown that a method can be uniquely reduced to and thus uniquely
identified by a small minset over \Lex and \LTT.  However, \LTT is too coarse.
Token types as are too abstract.  \Lex preserves lexical differences.  The
existence of minsets over \Lex can be of practical use in applications like Code
Search.  The problem with minsets over \Lex is that they do not capture behavior
and behavior differences well.  Raw lexemes are too specific and cryptic.  

\boldpara{Goal} We set the goal of finding a lexicon that is \emph{minimal}, as
small as possible, and \emph{natural}, consisting of words that a human would
know and use in applications like code search and code synthesis.  The words in
this lexicon should be \emph{meaningful}, in the sense that they reveal
information about the behavior of the code to us, humans.  Since a \Minset is,
by definition, distinguishing, we then expect minsets to capture behavior and
behavior differences of a piece of code.

\boldpara{Strategy} 
We search by exploring the lexicon spectrum toward more abstract views of code.
We additively construct a bag of words that approximates what a programmer might
naturally use in applications like code search and code synthesis.

\boldpara{Challenges} 
Two issues confounds this search: lexicon specialization can \emph{overfit} while
lexicon abstraction introduces \emph{imprecision}.  To ameliorate overfitting, we
restricted our search to natural lexicons.  By natural, we mean simple and
intuitive.  We pursue natural abstractions to avoid unnatural abstractions that
overfit our corpus, like one that maps every function in our corpus to a unique
meaningless word.  In our context, imprecision leads spurious homonyms which
reduces yield\footnote{\label{foot:syn}Although \Lex is rife with synonyms, our
candidate lexicons have almost none.}. To handle this problem, we relax the
definition of distinguishability (Definition \autoref{defn:lkd}). 
Henceforth, when we say distinguishable we mean $10$-distinguishable.  We chose
$10$ because that is consistent with what humans can process in a glance or two.
Humans can rapidly process short lists~\cite{miller1956magical}.
\begin{defn}
\label{defn:lkd} 
A unit of code is \emph{lexically k-distinguishable} if it does
not have distinguishing subset but has $10$ or fewer supersets.  
\end{defn}

We considered four candidates, lexicons. We listed and introduced them briefly
in \autoref{tab:lexicons}.  Our results appear in \autoref{fig:min50-min14} and
\autoref{fig:min50-min14-yield}.  We report absolute minset sizes.  In searching
or synthesizing code using minsets, the minset size is likely more important to
the programmer than the minset ratio.  We also report \emph{yield}, the
proportion of distinguishable methods.  The yield approximates the likelihood of
success for the programmer given that lexicon in the context of some code search
or code synthesis application, Broadly, it gives us a sense of the potential
practical usefulness of a lexicon.

\boldpara{\MinOne}
First, we considered \MinOne, a lexicon including only method names and
operators.  For public API methods, we used fully qualified method names to
prevent the spurious creation of homonyms.  For local methods, we abstracted all
names to a single abstract word to capture their presence.  Local methods tend
to implement project-specific functionality not provided by the public API, and
are not generally aimed for general use.  The intuition in including method
names is that a lot of the semantics is captured in method calls.  They are the
verbs or action words of program sentences.  Our intuition is further supported
by the effectiveness of API birthmarking~\cite{schuler2007dynamic}.  We also
included operators because all primitive program semantics are applications of
operators.  Using this lexicon, the mean and maximum minset sizes are small,
$2.73$ and $7$, respectively.  The imprecision of \MinOne
manifests itself in the low yield of $26.86$\%.


\boldpara{\MinTwo}
To try to improve yield, we created lexicon \MinTwo by including control flow
keywords as well; there are $13$ in Java.  From the programmer's perspective,
these words reveal a great deal about the structure of a method that is critical
to semantics.  For example, the word \lstinline{for} alone immediately tells us
that some behavior is repeated.  Using this lexicon, the mean and maximum minset
sizes are still small, $2.88$ and $9$, respectively.  The yield does not
increase much.  Only an additional $288$ methods become threshable. The likeliest
and simplest explanation for the small change is that these words are very
common; at least one of them is present in $83.26$\% of the methods.  It is more
difficult to interpret this change.  On the one hand, it is small.  On the other
hand, it is the result of adding only $13$ new, semantically-rich words.  In
balancing the size of lexicon with the interpretability of minsets, this appears
to be a good trade-off.


\boldpara{\MinThree}
In our quest to improve yield, we defined \MinThree to include the types of
variable identifiers (names).  Those of a public type were mapped to their fully
qualified type name.  Those of a locally-defined type were mapped to a single
abstract word to signal their presence.  Locally-defined types, like local
methods, tend to be project-specific and not of general use.   Our reason for
focusing on types is that they tell the programmer the kind of data on which
methods and operators act.  It is also a simple way of considering variable
identifiers.  Again, the mean and maximum minset size are small, $2.96$ and $9$,
respectively.  There is a notable increase in the yield, from $29.72$\% to
$41.44$\%.  It is now close to what we would imagine might be practical.  In a
\Minset-based application, a programmer would succeed $4$ out of $10$ times.
Though, the lexicon grew substantially by $36,260$ words.  This trade-off
appears reasonable considering as well that it is natural to supply the
programmer with the convenience of a variety of primitive and composite types.


\boldpara{\MinFour}
We defined a final lexicon, \MinFour, which includes \lstinline+false+,
\lstinline+true+, and \lstinline+null+, object reference keywords, like
\lstinline+this+ and \lstinline+new+, and the token types of constant values,
such as the token type \lstinline{Character-Literal} for `Z' or, for
\lstinline+5+, \lstinline+Integer-Literal+.  In total, we added $13$ new words.
Our intuition is that the \emph{use} of hard-coded strings and numbers is
connected to behavior.  Certainly, knowing that hard-coded values are used can
be informative.  Also, in an application, a programmer may need to indicate that
some constant string or number will be used.  For example, if the programmer
wishes to find a method that calculates the area of a circle, then it would be
natural to indicate that target method likely contains a float literal like
\lstinline+3.14+. After including these words, the mean and maximum minset size
remain small, $3.06$ and $10$, respectively.  The yield increased from $41.44$\%
to $44.79$\%.  Adding this small number of semantically-rich words to the
lexicon seems to be another reasonable exchange for a noticeable gain in yield:
under this lexicon, the words are easier to interpret (see
\autoref{sec:results:semantics} for our analysis of the interpretability of
minsets built from these words) while remaining small enough for humans to work
with, \eg a human could potentially write a minset from scratch while
programming using key words~\cite{little_ase_07}.

\subsection{The Effect of Multiplicity and Abnormally Large Methods on Distinguishability}
\label{sec:results:yield}

\begin{figure}[t]
  \centering
  \includegraphics[width=0.48\textwidth]{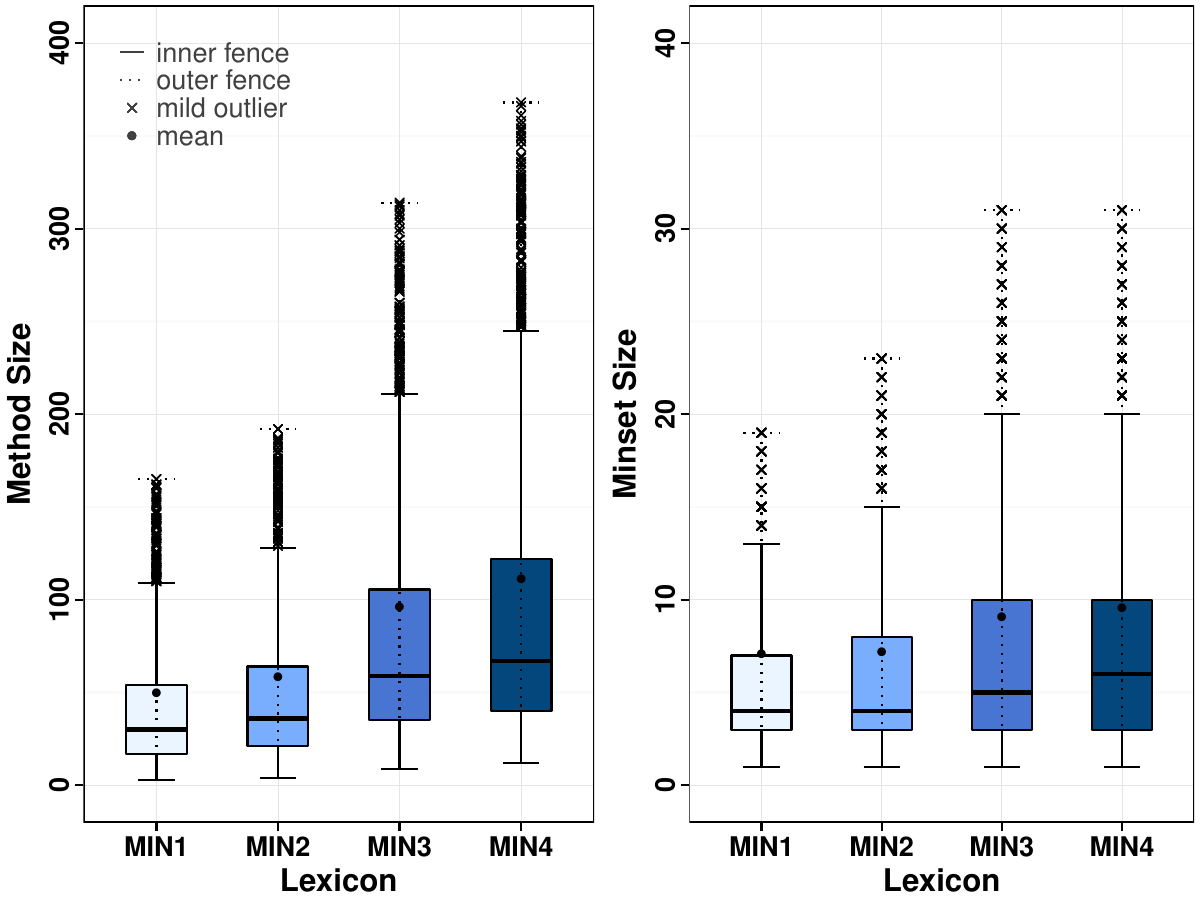}  \\
	\caption{Multiplicity: (left) Like in \autoref{fig:min50-min14}, as the
lexicon grows, so does the threshed method size.  In this case, methods are much
larger because repetition is allowed.  (right) The minset sizes, allowing
repetition, are evidently larger.  However, on average, they are still small
across all lexicons.  (To visualize both distributions, we omitted extreme
outliers.\protect\footnotemark \@)}

  \label{fig:min50-min14-m}
\end{figure}
\footnotetext{A point is an extreme outlier if it lies beyond $Q3 + 3*IQ$ or
below $Q1 - 3*IQ$, where $IQ = Q3 - Q1$.}

\begin{figure}[t]
  \centering
  \includegraphics[width=0.48\textwidth]{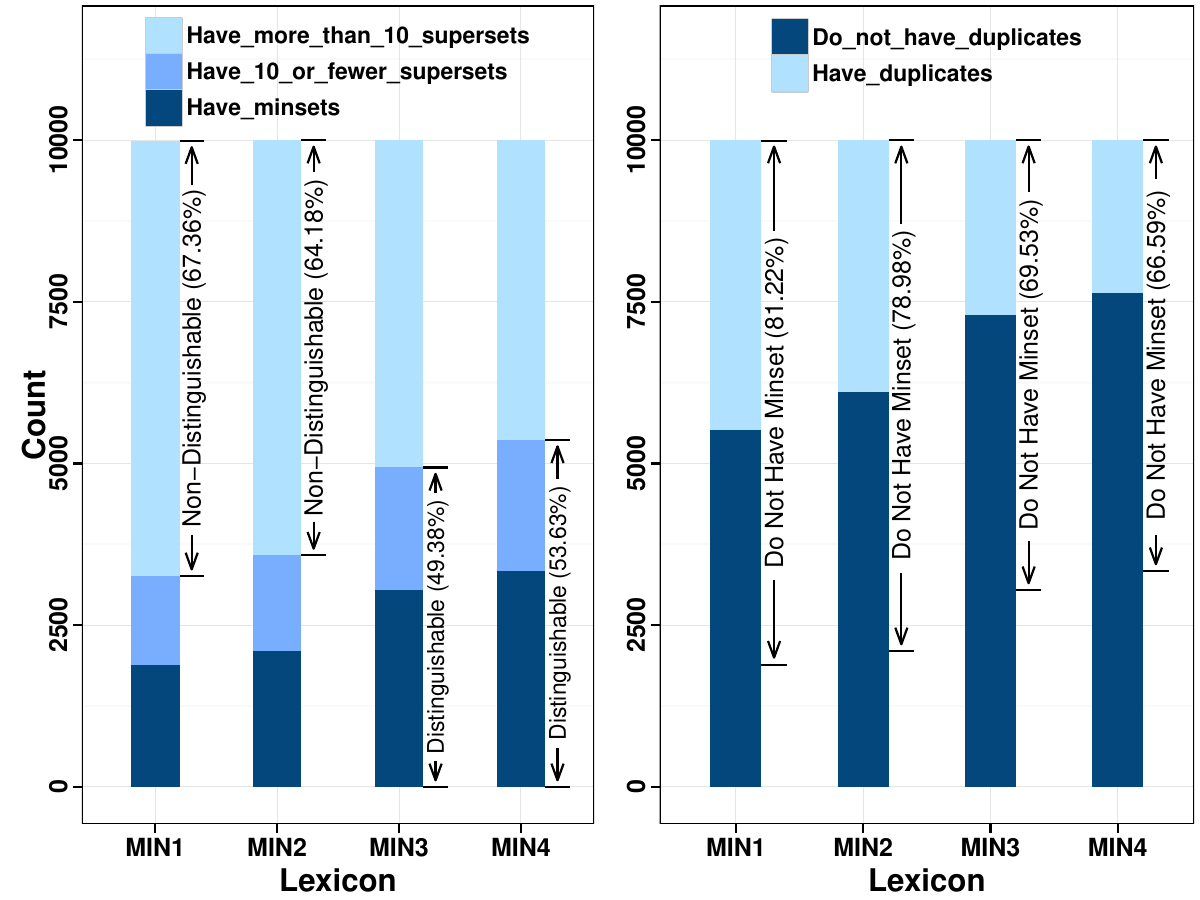}  \\
	\caption{Multiplicity: (left) \emph{Yield:} Multiplicity improves the yield of
all lexicons.  The yield of \MinFour now exceeds $50$\%. (right)
\emph{Proportion of Methods With Duplicates:} Using this proportion as a rough
measure of threshing, multiplicity also improves the threshing precision of each
lexicon.  Less than 25\% of the methods have duplicates using \MinFour.  (Note:
Compare with \autoref{fig:min50-min14-yield}.)}
  \label{fig:min50-min14-m-yield}
\end{figure}

Instead of continuing our search for lexicons generated from ever more complex
abstractions over lexemes, we reconsidered \emph{multiplicity}, the number of
copies of a word in a method.  We hypothesized that modeling methods as
multisets would recapture some lexical differences, and thereby increase the
yield of the lexicons \MinOne through \MinFour.  We used the multiset version of
\algref{alg:minset} to recompute minsets, and show our results in
\autoref{fig:min50-min14-m}.


Multiplicity improved yield at the cost of larger absolute minset sizes.  The
yield increased for all lexicons.  The new yields ranged from
$32.64$\%--$53.63$\%.  The smallest increase in yield was using \MinOne
($3.18$\%) and the largest was using \MinFour ($8.84$\%).  More concretely,
using \MinFour, the number of distinguishable methods increased by $884$.
Multiplicity also improved the minset ratios over all lexicons.  For example,
using \MinFour, the mean minset ratio decreased from $15.47$\% to $5.35$\%.  The
cost of considering multiplicity, however, was an overall increase in minset
sizes;  the range of mean minset sizes shifted, $2.73$--$3.06$, shifted and got
a bit wider, $7.06$--$9.56$.  The outliers of minset sizes moved farther to the
right. Previously, they ranged from $7$--$10$ and now they range from
$258$--$438$.  The right tails have grown longer.  For example, using \MinFour,
$75.67$\% of the minsets have fewer than $10$ words.  Another cost of the gain
in yield was in minset computation where we observed an approximate slowdown
factor ranging from $4$ to $7$.  For example, computing multiset minsets using
\MinOne took $44$ hours instead of $6$.  In practice, the slowdown is much
better than \algref{alg:minset}'s complexity implies.  Overall, despite its
cost, modeling methods as multisets over \MinFour produces a yield with
practical value: it easily distinguishes more than half of the methods in our
sample set.

Multiplicity appears to also improve the how well the bag-of-words preserves
lexical differences.  Modeling a method as a bag-of-words can map two unique
methods to the same set or multiset.  When this happens, the \Minset algorithm
cannot distinguish them.  We can use the proportion of methods with duplicates
to gauge the precision of the bag-of-words model.  \Lex gave us a baseline of
$3.20$\%.  When we experimented with lexicons \MinOne through \MinFour and no
multiplicity, we observed the portion improved from $66.4$\% using \MinOne down
to $41.64$\% using \MinFour (\autoref{fig:min50-min14-m-yield}).  Multiplicity
cut those figures nearly in half.  For example, using \MinFour, the proportion
of methods with duplicates is only $23.59$\%.

The remaining portion of non-distinguishable methods is still intriguing.  There
are still $46.37$\% non-distinguishable methods, entirely subsumed by more than
$10$ other methods.  We certainly expected some methods to subsume others
because of their sheer size.  We also expected families of methods with similar
behavior where some subsume others.  However, given that methods are not that
small, containing, on average, $72.8$ words over \MinFour, and that the the
portion of methods with duplicates is small, we suspected another reason.  We
hypothesized that there are abnormally large methods subsuming a great number of
methods.

\begin{figure}[t]
  \centering
  \includegraphics[width=0.48\textwidth]{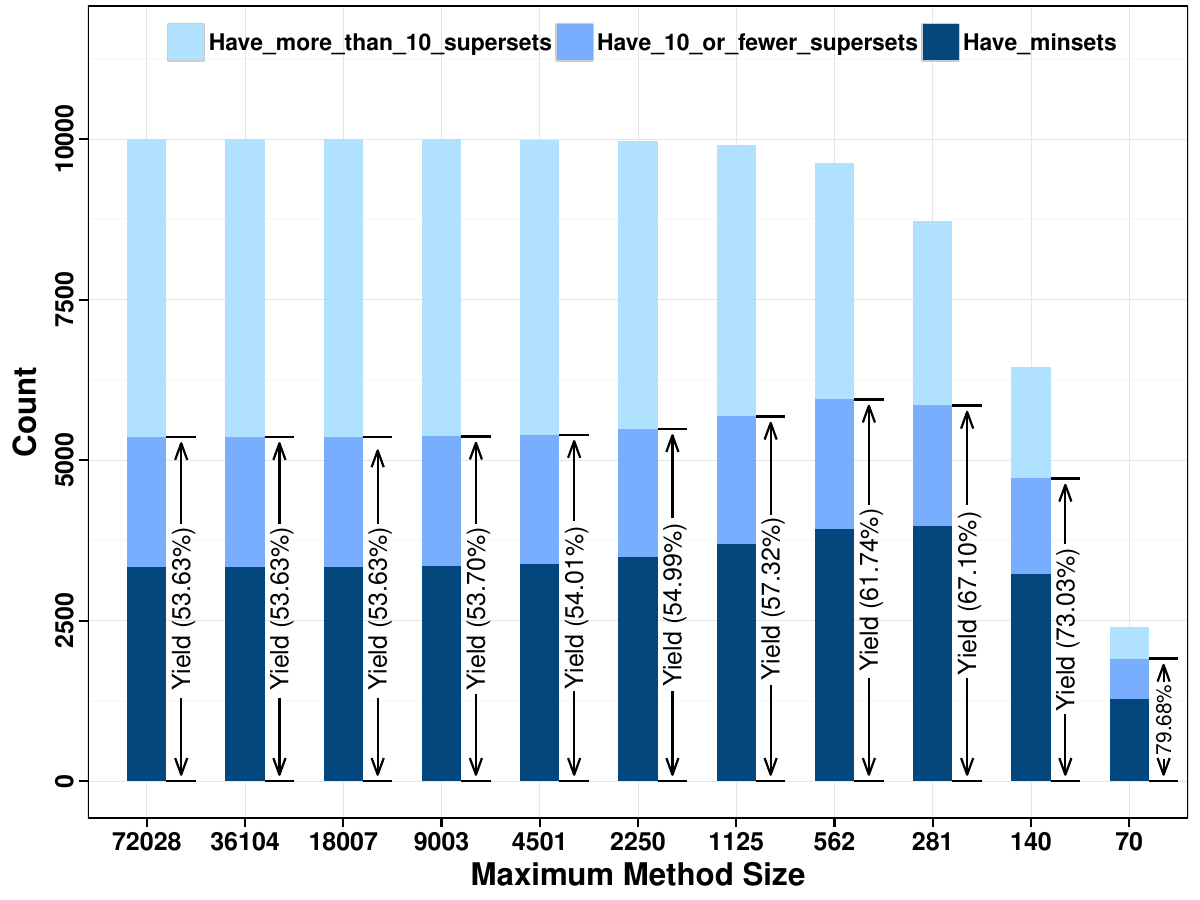}  \\
	\caption{Yield, the percentage of distinguishable methods, increases as the
maximum method size filter is tuned down to 562.  From there, the number of
methods and the number of threshable methods decreases substantially.  Thus,
setting the filter at 562 seems appropriate.}
	\label{fig:tunemaxsize}
\end{figure}

We conducted an experiment where we gradually filtered large methods to observe
the effect on yield (\autoref{fig:tunemaxsize}); we can perform this experiment
without recomputing minsets.  We initialized the filter size to $72,028$, the
maximum method size (in tokens) in our corpus, and repeatedly halved it down to
$70$; the miminum size of a method is $50$.  Yield increases as the maximum
method size filter is tuned down to $562$.  That appears to be the ``sweet
spot.'' If we filter methods with more than $562$ tokens, or about $56$ lines of
code, then the yield improves from $53.67$\% to $61.74$\%.  In an application
that implements this filter would means that, a user would succeed $6$ out of
$10$ times.  For example, in a code search application, the likelihood of
success of finding (recalling) a method would be improved if the application did
not consider abnormally large methods.  If we doubled the filter size to $1125$,
we would reconsider $55,953$ methods, and the yield would still be higher at
$57.32$\% than without the filter.  
Since there is a relatively low number of these large methods, $69,535$ out of
$1,870,905$ (or $3.7$\%), the trade-off seems reasonable.  A maximum size filter
would clearly add practical value to a \Minset-based application.

\MinFour is a natural lexicon suited for code search, synthesis, and robust
programming.  We recomputed minsets using \MinFour considering multiplicity and
a filter size $562$.  As we already mentioned, the yield is $61.74$\%.  The mean
minset size increases with the filter from $9.56$ to $11.03$.  The minset sizes
vary but have a clear positive skew where fewer than $25$\% contain more than
$12$ words.  That right tail of the distribution is significantly shorter; the
maximum size decreased from $689$ to $173$ because of the filter.

\subsection{Minset Over \MinFour}
\label{sec:results:semantics}
Minsets computed over \Lex are small but do not capture behavior well.  Minsets
over \MinFour are still small; a few words are needed to distinguish a unit of
code.  \emph{To what extent do minsets over \MinFour capture behavior and
behavior differences amongst methods?}  We provide a qualitative answer to this
question via case studies:  Minsets over \MinFour give insight into the behavior
of a method.

\begin{figure}[t]
  \centering
  \includegraphics[width=0.48\textwidth]{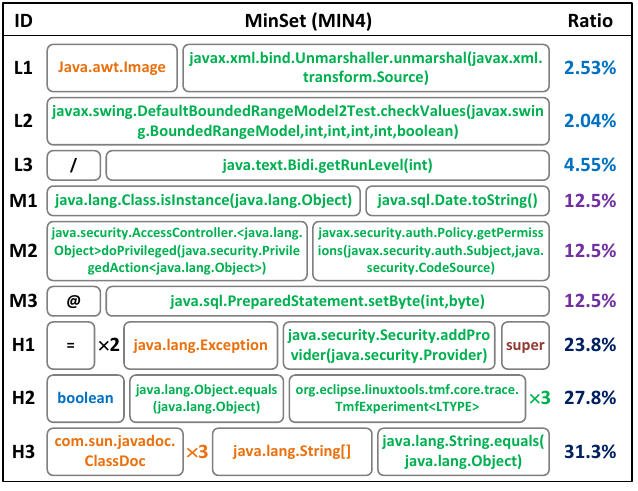}  \\
	\caption{This shows the minsets of nine methods (\MinFour).  L1-L3 are minsets
that have low minset ratios.  M1-M3 have medium minset ratios.  H1-H3 have high
minset ratios.  The minset elements are rich and reveal some information about
the behavior of their respective methods.}

	\label{fig:case-study}
\end{figure}

We studied the minsets produced in our last
experiment in \autoref{sec:results:yield}.  We selected nine minsets
(\autoref{fig:case-study}); we partitioned the methods into low, medium, and
high minset ratios and picked three uniformly at random from each subset.  For
each minset, we tried to understand each element and what they revealed together
about the behavior of a method.  Then we inspected the method source code more
carefully to assess how well the minsets capture method functionality.  Due to
lack of space, we discuss only three in detail.

\vspace*{6pt}
\noindent
\underline{\textit{Low}: \textbf{L1}}\quad The method named
\lstinline[literate={\\\-}{}{0\discretionary{-}{}{}}]+javax\-.\-xml\-.\-bind\-.\-Unmarshaller\-.\-unmarshal+
from
\lstinline[literate={\\\-}{}{0\discretionary{-}{}{}}]+(java\-.\-xml\-.\-transform\-.\-Source)+
deserializes XML documents and returns a Java content tree object;
\lstinline+java.awt.Image+ is an abstract classes that represents graphical images.
From this minset, we infer that this method handles images and \texttt{XML} files.
Since it reads the \texttt{XML} file, we also infer that it uses \texttt{XML} data
in some manner.
Perhaps the file contains a list of images, or the data in the file is used to create or alter an image.
After inspecting the source code, we find that it is a method in the
\lstinline+LargeInlineBinaryTestCases+ class of the Eclipse Link project,
which manages \texttt{XML} files and other data stores.
Our understanding was not far off: the method does read a binary \texttt{XML} file that contains images.

\vspace*{6pt}
\noindent
\underline{\textit{Medium}: \textbf{M1}}\quad 
The
\lstinline[literate={\\\-}{}{0\discretionary{-}{}{}}]+java.lang.Class.isInstance(java.lang.\-Object)+
method checks if a given object is an object of type \lstinline+Class+ or
assignment-compatible with its calling object.  The \lstinline+java.sql.Date.+
\lstinline+toString()+ method converts a \lstinline+Date+ object, which has been
wrapped as an \texttt{SQL} date value, to a \lstinline+String+.  From this
minset, we understand the type of a variable is checked.  Perhaps, reflection is
used on an object to ensure it is an instance of type \lstinline+Date+ before it
converted to a string, for printing or storage.  Inspecting the source code we
find that this method resides in the \lstinline+DateType+ class of the Hibernate
ORM project.  Again, our understanding is very close to the behavior of the
method.  The method is passed an object, which it ensures is a
\lstinline+java.sql.Date+ class object, and then returns the value as a string
in the appropriate \texttt{SQL} dialect.

\vspace*{6pt}
\noindent
\underline{\textit{High}: \textbf{H1}}\quad 
The \lstinline+java.lang.Exception+ object is thrown in Java to indicate abnormal
flow or behavior. The
\lstinline+=+ operator tells us that there is an assignment but is very common.
The
\lstinline[literate={\\\-}{}{0\discretionary{-}{}{}}]+java\-.\-security\-.\-Security\-.\-addProvider\-(\-java\-.\-security\-.\-Provider)+
method
adds a security service object, \lstinline+Provider+, to a
\lstinline+Security+ object.
The \lstinline+Security+ object centralizes all the security properties in an application.
The \lstinline+super+ keyword refers to the superclass.
From this minset, we can infer that it describes a constructor that probably 
overrides a method in its superclass.
We also infer that it catches an exception when adding the provider fails.
In the source, we confirm that it is a constructor in the
\lstinline+HsqlSocketFactorySecure+ class in the CloverETL project.
It wraps code that instantiates a \lstinline+Provider+ class and
adds it to the \lstinline+Security+ object in a \lstinline+try+ block.  If
adding the provider fails, it catches the exception, as we had inferred.

%% file: discussion.tex
\section{Discussion}
\label{sec:discussion}

Our results clearly support our Wheat and Chaff Hypothesis.  We have shown,
over a variety of lexicons, that functions are lexically distinguishable, and
that the distinguishing subsets tend to be small.  We defined and analyzed four
lexicons in search for a natural, minimal lexicon that induces more meaningful
minsets.  We offered \MinFour as the promising candidate. 

\boldpara{Other Lexicons} 
Our lexicon exploration avoided variable names because they are so
unconstrained, noisy, and rife with homonyms and synonyms.  Minsets over
lexicons, like \Lex, that incorporated them could include trivial, semantically
insignificant differences, like \lstinline+user+ \vs \lstinline+usr+ in Unix.
At the same time, variable names are an alluring source of signal.  Intuitively,
and in this corpus, they are the largest class of identifiers, which comprise
$70$\% of source code~\cite{deiss:iwpc:05}, and connect a program's source to
its problem domain~\cite{binkley:emse:13}.  In future work, we plan to separate
the ``wheat from the chaff'' in variable names.

\boldpara{Alternative Units of Code}
We chose functions as our unit of code.  However, we can apply the same
methodology at other syntactic levels.  One alternative is to study blocks of
code.  A single function can have many blocks.  This could be very useful in an
alternative programming model where the user seeks a common block of code but
for which there is no individual function.  Another alternative is to use
abstract syntax trees (AST) to preserve some syntactic structure in the lexical
features.  We could also consider using n-grams to preserve some order in the
features.
 

\boldpara{Threats to Validity}  
We identify two main threats.  The first is that we only studied Java.  However,
we have no reason to believe that the ``wheat and chaff'' hypothesis does not
hold for other programming languages.  Java, though more modern, was designed to
be very similar to C and C++ so that it could be adopted easily.  The second
threat comes from our corpus: size and diversity.  We downloaded a very large
corpus, by any standard.  In fact, we downloaded all the Java projects listed as
``Most Popular'' in the four code repositories we crawled.  Those code
repositories are known primarily for hosting open-source projects.  Thus, there
is no indication that they are biased toward any specific types of projects.  We
plan to replicate this study on a larger Java corpus and with languages of
different paradigms like Lisp and Prolog to help us understand to what extent
the lexical distinguishability phenomenon varies and to what extent the Wheat and
Chaff Hypothesis holds.

%% file: applications.tex
\section{Applications}
\label{sec:apps}

Though our study is primarily empirical, in this section, we describe
pre-existing and new applications for minsets.

\boldpara{SmartSynth (Existing)} 
As mentioned earlier, the clearest and, perhaps, most promising application for
minsets is in keyword-based programming.  SmartSynth~\cite{le_mobisys_13} is a
recent, modern incarnation.  SmartSynth generates a smartphone script from a
natural language description (query).  ``Speak weather in the morning'' is an
example of a successful query.  SmartSynth uses NLP techniques to parse the
query and map it to a set of ``components'' (words) in its underlying
programming language.  Combining a variety of techniques, it then infers
relationships between the words to generate and rank candidate scripts.  At its
heart is the idea that usable code can be constructed from a small set of words.
This subset is a minset or another distinguishing subset. 

\boldpara{Code Search Engine (New)} 
A major problem of code search is ranking
results~\cite{bajracharya2006oopsla,mandelin2005pldi,mcmillan2012icse}.  We
built a code search engine that uses a new ranking
scheme\footnote{\label{foot:ce}\scriptsize{\url{http://jarvis.cs.ucdavis.edu/code\_essence}}.}.
Relevant methods are ranked by the similarity between their minsets and the
user's query.  For example, the query ``sort array int'' returns $135$ methods.
The top result, with minset ``sort array parseInt 16'', returns a sorted array
of integers, if the `sort' flag is set. 

\boldpara{Code Summarizer (New)} 
From our case studies of \MinFour minsets, we realized that minsets can
effectively summarize code.  We built a code summary web
application\footnotemark[\getrefnumber{foot:ce}]. A user enters the source code
of a method, our tool computes a minset, and presents it as a concise summary.
Due to space constraints, we omit a full example and invite interested readers
to explore our web application.  \autoref{fig:case-study} shows examples of
minsets summarizing methods.

\boldpara{\Minset-powered IDE (Concept)}
Our results offer insight into how to develop a more powerful, alternative
programming system.  Consider an integrated development environment (IDE), like
Eclipse or IntelliJ, that can search a \Minset indexed database of code and
requirements to
\begin{inparaenum}[1)]
\item propose related code that may be adapted to purpose,
\item auto-complete whole code fragments as the programmer works,
\item speed concept location for navigation and debugging, and
\item support traceability by interconnecting requirements and
  code~\cite{cleland2005icse}.
\end{inparaenum}

%% file: relwork.tex
\section{Related Work}
\label{sec:relwork}

Although we are the first to study the phenomenon of lexical distinguishability
of source code, and propose the Wheat and Chaff Hypothesis\footnote{Others have
used the ``wheat and chaff'' analogy in the computing world but in different
domains~\cite{rivest1998online, schleimer2003sigmod}.}, a few strands of related
work exist.

\boldpara{Code Uniqueness} 
At a basic level, our study is about uniqueness.  \emph{What lexical features
distinguish or uniquely identify a unit of code?}  Gabel and Su also studied
uniqueness~\cite{gabel_fse_10}.  They found that software generally lacks
uniqueness which they measure as the proportion of unique, fixed-length token
sequences in a software project.  We studied uniqueness differently.  We capture
uniqueness as the size or proportion of minsets.  The elements in a \Minset may
not be unique or even rare but together uniquely identify a piece of code.  We
keep in mind that syntactic differences do not always imply functional
differences as Jiang and Su demonstrated~\cite{jiang_issta_09}.  Thus, in some
cases the uniqueness may be accidental.  Two minsets may, in fact, represent the
same behavior at some higher, more abstract semantic level.

\boldpara{Code Completion and Search} 
Observations about natural language phenomenon provide a promising path toward
making programming easier.  Hindle \etal focused on the ``naturalness'' of
software~\cite{hindle_icse_12}.  They showed that actual code is ``regular and
predictable'', like natural language utterances.  To do so, they trained an
$n$-gram model on part of a corpus, and then tested it on the rest.  They
leveraged code predictability to enhance Eclipse's code completion tool.  Their
work followed that of Gabel and Su who posited and gave supporting evidence that
we are approaching a ``singularity'', a point in time where all the small
fragments of code we need to write will already exist~\cite{gabel_fse_10}.  When
that happens, many programming tasks can be reduced to finding the desired code
in a corpus.  Our work suggests that small, natural set of words, \ie, minsets,
can index and retrieve code.  As for code completion, a \Minset-based approach
could exploit not just the previous $n-1$ tokens, but on all the previous tokens
and complete not just the next token but whole pieces of code.

Sourcerer and Portolio, two modern code search engines, support basic term
queries, in addition to more advanced
queries~\cite{bajracharya2006oopsla,mcmillan2011icse}.  Our research suggests
that the natural and efficient term query is a \Minset.  Search results may
differ in granularity.  Portfolio focuses on finding
functions~\cite{mcmillan2011icse} while Exemplar, another engine, finds whole
applications~\cite{grechanik2010icse}, \Minset easily generalizes to arbitrary
units of code.  Finally, code search must also be
``internet-scale''~\cite{gallardo2009suite}.  With a modest computer, we can
compute minsets for corpora of code of various languages, and update them
regularly as new code is added. 

Code completion tools suggest code a programmer \emph{might} want to use. They
infer relevant code and rank it.  Many diverse, useful tools and strategies
exist~\cite{bruch2009fse,nguyen2012icse,nguyen2013statistical,zhang2012icse}.
Our work suggests a different, complementary \Minset-based strategy: If what the
programmer is coding contains the \Minset of some existing piece of code,
suggest that. 

\boldpara{Genetics and Debugging}  
At a high-level, \algref{alg:minset} isolates a minimal set of essential
elements.  Central to synthetic biology is the search for the `minimal genome',
the minimal set of genes essential to living
organisms~\cite{acevedo2013trendsgenet}~\cite{maniloff1996pnas}.  Delta
debugging is very similar in that it finds a minimal set of lines of code that
trigger a bug~\cite{cleve2000aadebug}.  Both approaches rely on an oracle who
defines what is ``essential'' whereas we define ``essentialness'' with respect
to other sets.

%% file: conc.tex
\section{Conclusion and Future Work}
\label{sec:concfw}

We imagine that code, to the human mind, is amorphous, and ask: ``If a
programmer were reading this code, what features would be semantically
important?'' and ``If a programmer were trying to write this piece of code, what
key ideas would the programmer communicate?'' A \Minset is our proposal of a
useful, formal definition of these key ideas as `wheat.' Our definition is
constructive, so a computer can compute Minsets to generate or retrieve an
intended piece of code.

We evaluated Minsets, over a large corpus of real-world Java programs, using
various, natural lexicons: the computed minsets are sufficiently small and
understandable for use in code search, code completion, and natural programming.

%% file: paper.bbl
\begin{thebibliography}{10}

\bibitem{acevedo2013trendsgenet}
C.~G. Acevedo-Rocha, G.~Fang, M.~Schmidt, D.~W. Ussery, and A.~Danchin.
\newblock From essential to persistent genes: a functional approach to
  constructing synthetic life.
\newblock {\em Trends in Genetics}, 29(5):273--279, 2013.

\bibitem{bajracharya2006oopsla}
S.~Bajracharya, T.~Ngo, E.~Linstead, Y.~Dou, P.~Rigor, P.~Baldi, and C.~Lopes.
\newblock Sourcerer: a search engine for open source code supporting
  structure-based search.
\newblock In {\em Companion to the 21st ACM SIGPLAN Symposium on
  Object-Oriented Programming Systems, Languages, and Applications}, pages
  681--682, 2006.

\bibitem{basit2007tokenclonedetection}
H.~A. Basit and S.~Jarzabek.
\newblock Efficient token based clone detection with flexible tokenization.
\newblock In {\em Proceedings of the 6th Joint Meeting of the European Software
  Engineering Conference and the ACM SIGSOFT Symposium on the Foundations of
  Software Engineering}, pages 513--516, 2007.

\bibitem{binkley:emse:13}
D.~Binkley, M.~Davis, D.~Lawrie, J.~I. Maletic, C.~Morrell, and B.~Sharif.
\newblock The impact of identifier style on effort and comprehension.
\newblock {\em Empirical Software Engineering}, 18(2):219--276, Apr. 2013.

\bibitem{bruch2009fse}
M.~Bruch, M.~Monperrus, and M.~Mezini.
\newblock Learning from examples to improve code completion systems.
\newblock In {\em Proceedings of the 7th Joint Meeting of the European Software
  Engineering Conference and the ACM SIGSOFT Symposium on the Foundations of
  Software Engineering}, pages 213--222, 2009.

\bibitem{cleland2005icse}
J.~Cleland-Huang, R.~Settimi, O.~BenKhadra, E.~Berezhanskaya, and S.~Christina.
\newblock Goal-centric traceability for managing non-functional requirements.
\newblock In {\em Proceedings of the International Conference on Software
  Engineering}, pages 362--371, 2005.

\bibitem{cleve2000aadebug}
H.~Cleve and A.~Zeller.
\newblock Finding failure causes through automated testing.
\newblock In {\em Proceedings of the Fourth International Workshop on Automated
  Debugging}, 2000.

\bibitem{deiss:iwpc:05}
F.~Dei{\ss}enb\"ock and M.~Pizka.
\newblock Concise and consistent naming.
\newblock In {\em Proceedings of the 13th International Workshop on Program
  Comprehension}, pages 97--106, 2005.

\bibitem{gabel_fse_10}
M.~Gabel and Z.~Su.
\newblock A study of the uniqueness of source code.
\newblock In {\em Proceedings of the 18th ACM SIGSOFT Symposium on the
  Foundations of Software Engineering}, pages 147--156, 2010.

\bibitem{gallardo2009suite}
R.~E. Gallardo-Valencia and S.~Elliott~Sim.
\newblock Internet-scale code search.
\newblock In {\em Proceedings of the 2009 ICSE Workshop on Search-Driven
  Development-Users, Infrastructure, Tools and Evaluation}, pages 49--52, 2009.

\bibitem{grechanik2010icse}
M.~Grechanik, C.~Fu, Q.~Xie, C.~McMillan, D.~Poshyvanyk, and C.~Cumby.
\newblock A search engine for finding highly relevant applications.
\newblock In {\em Proceedings of the ACM/IEEE International Conference on
  Software Engineering}, pages 475--484, 2010.

\bibitem{hindle_icse_12}
A.~Hindle, E.~T. Barr, Z.~Su, M.~Gabel, and P.~Devanbu.
\newblock On the naturalness of software.
\newblock In {\em Proceedings of the International Conference on Software
  Engineering}, pages 837--847, 2012.

\bibitem{jiang_issta_09}
L.~Jiang and Z.~Su.
\newblock Automatic mining of functionally equivalent code fragments via random
  testing.
\newblock In {\em Proceedings of the 18th International Symposium on Software
  Testing and Analysis}, pages 81--92, 2009.

\bibitem{le_mobisys_13}
V.~Le, S.~Gulwani, and Z.~Su.
\newblock {SmartSynth}: synthesizing smartphone automation scripts from natural
  language.
\newblock In {\em Proceeding of the 11th Annual International Conference on
  Mobile Systems, Applications, and Services}, pages 193--206, 2013.

\bibitem{li2004cpminer}
Z.~Li, S.~Lu, S.~Myagmar, and Y.~Zhou.
\newblock {CP-Miner}: a tool for finding copy-paste and related bugs in
  operating system code.
\newblock In {\em Proceedings of the Symposium on Operating Systems Design \&
  Implementation}, pages 289--302, 2004.

\bibitem{little_ase_07}
G.~Little and R.~C. Miller.
\newblock Keyword programming in {J}ava.
\newblock In {\em Proceedings of the IEEE/ACM International Conference on
  Automated Software Engineering}, pages 84--93, 2007.

\bibitem{little_nocode_10}
G.~Little, R.~C. Miller, V.~H. Chou, M.~Bernstein, T.~Lau, and A.~Cypher.
\newblock Sloppy programming.
\newblock In A.~Cypher, M.~Dontcheva, T.~Lau, and J.~Nichols, editors, {\em No
  Code Required}, pages 289--307. Morgan Kaufmann, 2010.

\bibitem{mandelin2005pldi}
D.~Mandelin, L.~Xu, R.~Bod\'{\i}k, and D.~Kimelman.
\newblock Jungloid mining: helping to navigate the {API} jungle.
\newblock In {\em Proceedings of the 2005 ACM SIGPLAN Conference on Programming
  Language Design and Implementation}, pages 48--61, 2005.

\bibitem{maniloff1996pnas}
J.~Maniloff.
\newblock The minimal cell genome: "on being the right size".
\newblock {\em Proceedings of the National Academy of Sciences},
  93(19):10004--10006, 1996.

\bibitem{mcmillan2011icse}
C.~McMillan, M.~Grechanik, D.~Poshyvanyk, Q.~Xie, and C.~Fu.
\newblock Portfolio: finding relevant functions and their usage.
\newblock In {\em Proceedings of the 33rd International Conference on Software
  Engineering}, pages 111--120, 2011.

\bibitem{mcmillan2012icse}
C.~McMillan, N.~Hariri, D.~Poshyvanyk, J.~Cleland-Huang, and B.~Mobasher.
\newblock Recommending source code for use in rapid software prototypes.
\newblock In {\em Proceedings of the 34th International Conference on Software
  Engineering}, pages 848--858, 2012.

\bibitem{miller1956magical}
G.~A. Miller.
\newblock The magical number seven, plus or minus two: some limits on our
  capacity for processing information.
\newblock {\em Psychological review}, 63(2):81, 1956.

\bibitem{miller_uist_08}
R.~C. Miller, V.~H. Chou, M.~Bernstein, G.~Little, M.~Van~Kleek, D.~Karger, and
  m.~schraefel.
\newblock Inky: a sloppy command line for the web with rich visual feedback.
\newblock In {\em Proceedings of the 21st Annual ACM Symposium on User
  Interface Software and Technology}, pages 131--140, 2008.

\bibitem{nguyen2012icse}
A.~T. Nguyen, T.~T. Nguyen, H.~A. Nguyen, A.~Tamrawi, H.~V. Nguyen,
  J.~Al-Kofahi, and T.~N. Nguyen.
\newblock Graph-based pattern-oriented, context-sensitive source code
  completion.
\newblock In {\em Proceedings of the 34th International Conference on Software
  Engineering}, pages 69--79, 2012.

\bibitem{nguyen2013statistical}
T.~T. Nguyen, A.~T. Nguyen, H.~A. Nguyen, and T.~N. Nguyen.
\newblock A statistical semantic language model for source code.
\newblock In {\em Proceedings of the 9th Joint Meeting of the European Software
  Engineering Conference and the ACM SIGSOFT Symposium on the Foundations of
  Software Engineering}, 2013.

\bibitem{openjdk_12}
Oracle open{JDK}.
\newblock \url{http://openjdk.java.net/}, 2012.

\bibitem{reiss2009icse}
S.~P. Reiss.
\newblock Semantics-based code search.
\newblock In {\em Proceedings of the 31st International Conference on Software
  Engineering}, pages 243--253, 2009.

\bibitem{reiss2009suite}
S.~P. Reiss.
\newblock Specifying what to search for.
\newblock In {\em Proceedings of the 2009 ICSE Workshop on Search-Driven
  Development-Users, Infrastructure, Tools and Evaluation}, pages 41--44, 2009.

\bibitem{rivest1998online}
R.~Rivest.
\newblock Chaffing and winnowing: Confidentiality without encryption, March
  1998.
\newblock web page.

\bibitem{schleimer2003sigmod}
S.~Schleimer, D.~S. Wilkerson, and A.~Aiken.
\newblock Winnowing: Local algorithms for document fingerprinting.
\newblock In {\em Proceedings of the 2003 ACM SIGMOD International Conference
  on Management of Data}, SIGMOD '03, pages 76--85, New York, NY, USA, 2003.
  ACM.

\bibitem{schuler2007dynamic}
D.~Schuler, V.~Dallmeier, and C.~Lindig.
\newblock A dynamic birthmark for {J}ava.
\newblock In {\em Proceedings of the International Conference on Automated
  Software Engineering}, pages 274--283, 2007.

\bibitem{zhang2012icse}
C.~Zhang, J.~Yang, Y.~Zhang, J.~Fan, X.~Zhang, J.~Zhao, and P.~Ou.
\newblock Automatic parameter recommendation for practical {API} usage.
\newblock In {\em Proceedings of the 34th International Conference on Software
  Engineering}, pages 826--836, 2012.

\end{thebibliography}
